\documentclass[12pt]{article}
\usepackage{amsmath,amssymb,bm,epsfig}

\renewcommand{\thefootnote}{\fnsymbol{footnote}}

\begin{document}

\title{
\begin{flushright}
\begin{minipage}{0.2\linewidth}
\normalsize
WU-HEP-16-06 \\
EPHOU-16-001 \\*[50pt]
\end{minipage}
\end{flushright}
{\Large \bf 
Moduli mediation without moduli-induced gravitino problem
\\*[20pt]}}

\author{
Kensuke Akita$^{1,}$\footnote{
E-mail address: ken8a1@asagi.waseda.jp},\ \ 
Tatsuo~Kobayashi$^{2}$\footnote{
E-mail address:  kobayashi@particle.sci.hokudai.ac.jp}, \ \
Akane Oikawa$^{1,}$\footnote{
E-mail address:  a.oikawa@aoni.waseda.jp},\ and \ 
Hajime~Otsuka$^{1,}$\footnote{
E-mail address: h.otsuka@aoni.waseda.jp
}\\*[20pt]
$^1${\it \normalsize 
Department of Physics, Waseda University, 
Tokyo 169-8555, Japan} \\
$^2${\it \normalsize 
Department of Physics, Hokkaido University, Sapporo 060-0810, Japan} 
\\*[50pt]
}

\date{
\centerline{\small \bf Abstract}
\begin{minipage}{0.9\linewidth}
\medskip 
\medskip 
\small
We study the moduli-induced gravitino problem within the framework of the 
phenomenologically attractive mirage mediations. 
The huge amount of gravitino generated by the moduli decay 
can be successfully diluted by introducing an extra light modulus field which 
does not induce the supersymmetry breaking. 
Since the lifetime of extra  modulus field becomes longer than 
usually considered modulus field, our proposed mechanism is applied to both the 
low- and high-scale supersymmetry breaking scenarios. 
We also point out that such an extra modulus field appears in the flux compactification of 
type II string theory. 
\end{minipage}
}

\begin{titlepage}
\maketitle
\thispagestyle{empty}
\clearpage
\tableofcontents
\thispagestyle{empty}
\end{titlepage}

\renewcommand{\thefootnote}{\arabic{footnote}}
\setcounter{footnote}{0}
\vspace{35pt}

\section{Introduction}
Supersymmetry (SUSY) is a not only a phenomenologically plausible 
symmetry beyond the standard model (SM), but also is expected to 
appear in the low-energy effective theory of superstring theory. 
However, the lack of evidence of supersymmetry particles implies that 
the SUSY is broken above the TeV scale. 
In order to build in the SUSY-breaking scenario, 
we have to take care of the constraints from the collider experiments 
and cosmological observations, simultaneously. 
In particular, the mirage mediation~\cite{Choi:2004sx,Choi:2005ge,Choi:2005uz,Endo:2005uy}, 
which is the mixture of modulus~\cite{Kaplunovsky:1993rd,Brignole:1993dj,Kobayashi:1994eh,Ibanez:1998rf} and anomaly mediations~\cite{Randall:1998uk,Giudice:1998xp}, 
predicts the characteristic sparticle spectrum in contrast to the other 
SUSY-breaking scenarios. 
The modulus mediation is achieved by the framework of Kachru-Kallosh-Linde-Trivedi (KKLT)-type 
moduli stabilization~\cite{Kachru:2003aw}, where the volume modulus $T$ is stabilized at the 
AdS vacuum by the non-perturbative effects such as gaugino condensation 
on hidden D$7$-branes and Euclidean D-brane instanton, and 
the AdS vacuum is lifted to dS vacuum by anti D-branes. 
Since the {\it F}-term of $T$ is one-loop suppressed, the modulus mediation 
is comparable to the anomaly mediation. 
In this setup, the anomaly mediation and renormalization group effects cancel 
each other at a certain energy scale \cite{Choi:2005uz}, 
and the pure modulus mediation appears at that energy scale.
Such an energy scale is called as the mirage scale.
In particular, the TeV scale mirage mediation is important, 
because one can relax the fine-tuning problem on the Higgs mass \cite{Choi:2005hd,Kitano:2005wc,Choi:2006xb}.
For example, in the next-to-minimal supersymmetric standard model with the TeV scale mirage mediation, 
one needs ${\cal O}(10) \%$ tuning for $1.5$ TeV gluino mass and  ${\cal O}(1) \%$ fine-tuning even for 
several TeV of gluino mass to realize the weak scale \cite{Kobayashi:2012ee,Hagimoto:2015tua}.\footnote{
See also Ref.~\cite{Asano:2012sv}.}

In this paper, we focus on the cosmological aspects of pure modulus and mirage mediations. 
In the inflationary regime, the moduli fields are generically 
stabilized at the minimum away from those of KKLT-type moduli stabilization. 
This is because the moduli fields would receive the Hubble-induced masses due to 
the positive vacuum energy density or  they couple to the inflaton field through the 
Planck-suppressed operators. 
Even if the moduli fields do not receive the Hubble-correction due to the 
shift symmetry, the quantum fluctuations  deviate the minimum of moduli 
fields during the inflation. 
In any cases, 
when the Hubble scale is comparable to the masses of moduli fields, the moduli fields 
would oscillate around the true minima and 
such oscillating energy density dominates the energy density of the Universe. 
This is problematic from a cosmological point of view. 
First of all, the moduli fields should 
decay into the light particles before the start of Big Bang 
Nucleosynthesis (BBN) not to spoil the success of BBN. 
In addition, the moduli fields produce the huge 
amount of gravitinos. 
When the gravitino is unstable, the non-thermal lightest supersymmetric 
particle (LSP) is overproduced by the gravitino decay~\cite{Endo:2006zj,Nakamura:2006uc,Dine:2006ii}. 
Since this moduli-induced gravitino problem occurs in both low-and high-scale SUSY-breaking 
scenarios, it motivates us to explore dilution of the gravitino abundance. 
There are several studies to dilute the overproduced LSP by the thermal inflation~\cite{Lyth:1995hj,Lyth:1995ka}, 
Q-ball~\cite{Fujii:2001xp} and unstable domain-wall~\cite{Hattori:2015xla}, or 
the introduction of the axion sector~\cite{Nakamura:2008ey}. 
The modulus oscillation may be suppressed by the adiabatic oscillation~\cite{Linde:1996cx,Nakayama:2011wqa}. 
Note that the heavier gravitino compared with the volume modulus 
is not relevant for the above moduli-induced gravitino problem, as can be shown 
in the large volume scenario~\cite{Balasubramanian:2005zx,Conlon:2005ki}. 

In this paper, we propose a new dilution mechanism by an inclusion of the 
extra light modulus field, which does not break the SUSY at the vacuum. 
In the framework of flux compactification of type IIB string theory on 
Calabi-Yau (CY) manifold, 
it was argued that on general grounds all the complex structure moduli and 
axion-dilaton are stabilized at the compactification scale~\cite{Giddings:2001yu}. 
However, it depends on the choice of three-form fluxes. 
We consider that one of the complex structure moduli remains massless 
under the flux compactification, and it can be stabilized by the instanton 
effects without breaking the SUSY. 
Since such complex structure modulus is lighter than K\"ahler modulus, 
it plays an important role of diluting the gravitino produced by the K\"ahler modulus. 
After briefly reviewing the cosmological aspects of mirage mediation, such as moduli-induced 
gravitino problem in Sec.~\ref{sec:review}, we study the dilution mechanism based on 
the $4$D effective ${\cal N}=1$ supergravity (SUGRA) in Sec.~\ref{sec:SUGRA} and 
the effective action of type II string theory in 
Sec.~\ref{sec:typeII}, respectively. 
Finally, we conclude in Sec.~\ref{sec:con}.

\section{Brief review of the moduli-induced gravitino problem in the mirage mediation}
\label{sec:review}
We start with the $4$D SUGRA originating from 
the type IIB string theory on CY orientifold. 
The moduli K\"ahler potential is described in the reduced Planck 
unit\footnote{In this paper, we work the reduced Planck unit $M_{\rm Pl}=2.4\times 10^{18}$ GeV, 
unless we specify it.},
\begin{align}
K&=-3\ln (-i(T-\bar{T})) -\ln (-i(\tau -\bar{\tau}))\nonumber\\
&-\ln 
\Bigl[ 2i(F-\bar{F}) -i(U^i-\bar{U}^i) (\partial_i F +\partial_{\bar{i}} \bar{F}) \Bigl], 
\label{eq:Kahler}
\end{align}
where $T$ is the simplified overall K\"ahler modulus, $\tau$ is the axion-dilaton, 
and $F$ is the prepotential as functions of complex structure moduli 
$U^i$ $(i=1,2,\cdots, h^{1,2})$ with $h^{1,2}$ being the hodge number of CY manifold. 
The background three-form fluxes allow us to stabilize all the complex structure 
moduli and axion-dilaton with the following superpotential~\cite{Gukov:1999ya},
\begin{align}
W&=\sum_{\alpha=1}^{2h^{1,2}+2}(f^\alpha -\tau h^\alpha) \Pi_\alpha,
\label{eq:fluxW}
\end{align}
where $f^\alpha$ ($h^\alpha$) denotes the quanta of three-form fluxes relevant for the 
Ramond (Neveu-Schwarz) sector, and $\Pi_\alpha$ is the period vector 
of CY manifold determined by the prepotential $F$. 
Below the mass scales of stabilized complex structure moduli and axion-dilaton, 
we can extract the effective potential of K\"ahler modulus. 
Within the framework of KKLT moduli stabilization, the K\"ahler potential and 
superpotential for overall K\"ahler modulus $T$ read as
\begin{align}
K&=-3\ln (-i(T-\bar{T})) +{\rm const.},\nonumber\\
W&=w_0 +Ae^{iaT},
\label{eq:KW}
\end{align}
where $w_0$ is the real constant determined by the vacuum expectation values 
of the stabilized moduli, 
and the second term of superpotential is genereted by non-perturbative effects, 
e.g., gaugino condensation on $SU(N)$ pure super Yang-Mills 
theory living on hidden D$7$-brane with 
$a=8\pi^2/N$ and $A$ being real constant. 
By representing the chiral superfields as their scalar components in the same 
notation, the stabilization condition of overall volume modulus, 
$T=T_{\cal R}+iT_{\cal I}$, is given by
\begin{align}
&\langle T_{\cal I}\rangle \sim \frac{1}{a}\ln\left(-\frac{2A}{3w_0}\right),\nonumber\\
&\langle T_{\cal R}\rangle= 0,
\label{eq:TRI}
\end{align}
which leads to the AdS vacuum. 
After uplifting the AdS vacuum to dS vacuum, the K\"ahler modulus obtains 
the {\it F}-term, $F^T=-e^{K/2}K^{T\bar{T}}(\partial_{\bar{T}} \bar{W} +K_{\bar{T}}\bar{W})$ 
with $K^{T\bar{T}}$ being the inverse of K\"ahler metric $K_{T\bar{T}}=\partial_T\partial_{\bar{T}}K$, 
written in the gravitino mass $m_{3/2}=e^{\langle K/2\rangle} \langle W\rangle$, 
\begin{align}
\left\langle\frac{F^T}{T-\bar{T}}\right\rangle
 \simeq \frac{3m_{3/2}}{2aT_{\cal I}},
\end{align}
which is comparable to the anomaly mediation~\cite{Randall:1998uk,Giudice:1998xp} 
in the case of $a\langle T_{\cal I}\rangle ={\cal O}(4\pi^2)$,
\begin{align}
\frac{1}{16\pi^2}\left\langle\frac{F^C}{C_0}\right\rangle \simeq \frac{m_{3/2}}{16\pi^2} 
+\frac{(\partial_T K)F^T}{16\pi^2},
\end{align}
where $C=C_0+\theta^2 F^C$ denotes the chiral compensator. 
When the K\"ahler modulus is stabilized at the racetrack minimum~\cite{Krasnikov:1987jj,Dixon:1990ds} or 
it couples to the SUSY-breaking field in the {\it F}-term uplifting 
scenario~\cite{Lebedev:2006qq,Dudas:2006gr,Abe:2006xp,Kallosh:2006dv,Abe:2007yb}, 
it is possible to change the ratio of anomaly to modulus mediation.
(See for more details, e.g., Refs.~\cite{Dudas:2006gr,Abe:2006xp,Kallosh:2006dv,Abe:2007yb}.)
This results in the characteristic pattern of supersymmetric spectra without requiring the 
severe tuning. 

The mirage mediation is theoretically and phenomenologically 
attractive scenario.
Here, let us revisit  cosmological aspects of the mirage mediation. 
During the inflationary era, 
the moduli fields generically stay at the minimum away from that of 
KKLT-type moduli stabilization by the Hubble-induced mass. 
Thus, when the Hubble scale becomes comparable to the supersymmetric 
modulus mass, the modulus field rolls down to its true minimum and 
oscillates around it. 
Such oscillating energy  dominates the energy density of the Universe. 
The total decay rate of modulus field is mostly captured by the 
modulus decay into gauge boson and gaugino pairs. 
The relevant Lagrangian density is described by
\begin{align}
{\cal L}=\int d^2\theta \left[ \sum_{a=1}^3 \frac{f_a}{4} W^{a\alpha}W_{\alpha}^a +{\rm h.c.}\right],
\end{align}
where $W_{a\alpha}$ are the gauge field strength superfields with $a=1,2,3$ 
being the standard model gauge group, $U(1)_Y$, $SU(2)_L$, $SU(3)_C$, and 
$f_a$ are the gauge kinetic functions. 
When the standard model gauge group is derived from a single stack of 
D$7$-branes, the gauge kinetic function is given by
\begin{align}
f_a=k T +\Delta f_a (\langle U\rangle, \langle S\rangle),
\end{align}
where $k$ is constant and $\Delta f_a$ are heavy moduli-dependent 
threshold corrections. 
The decay width of real and imaginary parts of moduli into gauge boson and gaugino 
pairs are the same as each other, 
\begin{align}
\Gamma_{\rm tot}^T 
=\sum_{a=1}^3 \frac{N^a}{64\pi} \left(\frac{|k|\,g_a^2}{\sqrt{\langle K_{T\bar{T}}\rangle }}\right)^2 
\frac{m_T^3}{M_{\rm Pl}^2}
\simeq  2.9\times 10^{-21} |k|^2\langle K^{T\bar{T}}\rangle\left( \frac{m_T}{10^6\,{\rm GeV}}\right)^{3}\,{\rm GeV},
\label{eq:Ttot}
\end{align}
with $N^a=\{8, 3, 1\}$ and $(g_a)^{2}\simeq 0.53$ be the gauge couplings 
at the grand unification scale in the case of minimal supersymmetric standard model (MSSM), 
which gives rise to a radiation-dominated Universe with a 
reheating temperature,
\begin{align}
T_{\rm reh}^T\simeq \left(\frac{\pi^2g_\ast (T_{\rm reh}^T)}{90}\right)^{-1/4} 
\sqrt{ M_{\rm Pl}\Gamma_{\rm tot}^T} 
\simeq  80\,|k|\,\langle K^{T\bar{T}}\rangle^{1/2} \left( \frac{m_T}{10^6\,{\rm GeV}}\right)^{3/2}\,{\rm MeV}.
\end{align}
Here, $g_\ast (T_{\rm reh}^T) \simeq 10.75$ represents the effective number of degrees of freedom 
at the temperature $T_{\rm reh}^T$. 

In addition, the gravitino is also generated by the modulus decay with the following 
Lagrangian density in the unitary gauge,
\begin{align}
{\cal L}_{3/2} &= -\cfrac{\epsilon^{\mu\nu\rho\sigma}}{2} {\bar \Psi}_\mu \gamma_5 \gamma_\nu \partial_\rho \Psi_\sigma 
+\cfrac{\epsilon^{\mu\nu\rho\sigma}}{8} \left(\langle G_{T}\rangle \partial_\rho \tilde{T} -\langle G_{\overline{T}}\rangle \partial_\rho \overline{\tilde{T}}\right){\bar \Psi}_\mu \gamma_\nu \Psi_\sigma \nonumber\\
&\hspace{13pt}-\cfrac{1}{4}m_{3/2} {\bar \Psi}_\mu [\gamma^\mu, \gamma^\nu] \Psi_\nu 
-\cfrac{1}{8}m_{3/2} \left(\langle G_{T}\rangle \tilde{T} +\langle G_{{\overline T}}\rangle \overline{\tilde{T}}\right) {\bar \Psi}_\mu [\gamma^\mu, \gamma^\nu] \Psi_\nu,   
\label{eq:gravitinocou}
\end{align}
where $\Psi_\mu$ denotes the gravitino in the four-component formalism, $\tilde{T}=T-\langle T\rangle$, and 
$G_T=\partial G/\partial T$ with $G=K+\ln |W|^2$. 
We find that the decay width from the canonically normalized inflaton into the gravitino pair is 
\begin{align}
\Gamma_{3/2}^T&\simeq \frac{1}{288\pi \langle K_{T \bar{T}}\rangle} \left|\left\langle \frac{D_{T}W}{W} \right\rangle \right|^2 \cfrac{m_{T}^5}{m_{3/2}^2 
M_{\rm Pl}^2} \nonumber\\
&= \frac{|\langle K^{T\bar{T}}\rangle|^2}{96\pi } \left\langle\frac{F^T}{T-\bar{T}}\right\rangle^2 \cfrac{m_{T}^5}{m_{3/2}^4 
M_{\rm Pl}^2},
\end{align}
with $D_TW=\partial_{T}W +(\partial_TK)W$.

Let us denote the branching ratio from modulus to gravitino as $B_{3/2}$ 
which is typically of ${\cal O}(0.01-0.1)$ as pointed out in Ref.~\cite{Endo:2006zj}. 
Then, the gravitino yield which is the ratio of number density of gravitino to entropy density of 
the Universe, is estimated as
\begin{align}
Y_{3/2} \equiv \frac{n_{3/2}}{s} \simeq 2B_{3/2}\frac{3T_{\rm reh}^T}{4m_{T}} 
\simeq 
1.2\times 10^{-7} B_{3/2}|k|\,\langle K^{T\bar{T}}\rangle^{1/2}
\left( \frac{m_T}{10^6\,{\rm GeV}}\right)^{1/2},
\label{eq:Y32low}
\end{align}
which is preserved until the gravitino decays. 
When the non-thermally produced gravitinos decay into all the MSSM particles 
with the decay width $\Gamma_{3/2}=193\,m_{3/2}^3/(384\pi M_{\rm Pl}^2)$, 
the decay temperature of gravitino becomes
\begin{align}
T_{3/2} &\simeq \left(\frac{90}{\pi^2g_\ast (T_{3/2})}\right)^{1/4} \sqrt{\Gamma_{3/2}M_{\rm Pl}} 
\nonumber\\
&\simeq 7.8\left(\frac{10.75}{g_\ast (T_{3/2})}\right)^{1/4}
\left(\frac{m_{3/2}}{10^{5}\,{\rm GeV}}\right)^{3/2}\,{\rm MeV}. 
\end{align}
Note that the gravitino does not dominate the Universe at the  decay of 
gravitino, since the energy density of the gravitino $\rho_{3/2}$ is not larger than 
that of radiation $\rho_{r}$,
\begin{align}
\frac{\rho_{3/2}}{\rho_r}\simeq 
\left\{
\begin{array}{c}
\frac{T_{\rm NR}}{T_{3/2}}B_{3/2} 
<1\,\,\,\,(T_{\rm NR}>T_{3/2}),
\nonumber\\
B_{3/2} 
<1 \,\,\,\,\,(T_{\rm NR}<T_{3/2}),
\end{array}
\right.
\end{align}
where we denote the nonrelativistic temperature of gravitino as 
$T_{\rm NR}\simeq (m_{3/2}/m_{T})T_{\rm reh}^T$.

Finally, the dark matter is generated from the gravitino decay. 
Here and in what follows, we assume that the dark matter is consisted of 
the LSP under the assumption of $R$-pality conservation. 
The relic abundance of the dark matter, $Y_{\rm LSP}$, is found by solving 
the Boltzmann equation~\cite{Moroi:1999zb}, 
\begin{align}
Y_{\rm LSP}^{-1} =Y_{3/2}^{-1}\biggl|_{T=T_{3/2}} +\left(\sqrt{\frac{45}{8\pi^2g_\ast (T_{3/2})}} \frac{1}{M_{\rm Pl}T_{3/2}\langle \sigma_{\rm ann}v\rangle}\right)^{-1}\Biggl|_{T=T_{3/2}}, 
\label{eq:dmyield}
\end{align} 
where $T_{3/2}$ is the decay temperature of gravitino and $\langle \sigma_{\rm ann}v\rangle$ 
is the thermally averaged annihilation cross section of the dark matter. 

Now we take a closer look at the cosmological aspects of mirage mediation in the light of 
gravitino mass. 
When the gravitino mass is of ${\cal O}(30)$ TeV, 
the gravitino spoils the successful BBN unless $Y_{3/2}$ must be 
smaller than ${\cal O}(10^{-12})$~\cite{Kawasaki:2004qu,Kawasaki:2004yh,Kohri:2005wn}. 
Thus, Eq.~(\ref{eq:Y32low}) implies that the branching ratio, $B_{3/2}$, should be smaller than ${\cal O}(10^{-5})$. 
Even if the gravitino mass is larger than ${\cal O}(30)$ TeV, 
the dark matter abundance produced by the gravitino decay is overabundant 
to the Planck result as shown later. 
 
First we take up the low-scale SUSY-breaking scenario, where 
the mass of LSP is of ${\cal O}(100)$ GeV. 
When the annihilation of the LSPs produced by the gravitino is not effective, 
the first term in Eq.~(\ref{eq:dmyield}) dominates and 
the dark matter abundance is approximately given by 
\begin{align}
\Omega_{\rm LSP}h^2\simeq 
m_{\rm LSP} Y_{3/2} \frac{s_{\rm now}}{\rho_{\rm cr}} \simeq 
3342 B_{3/2}|k|\,\langle K^{T\bar{T}}\rangle^{1/2}
\left( \frac{m_{\rm LSP}}{100\,{\rm GeV}}\right)
\left( \frac{m_T}{10^6\,{\rm GeV}}\right)^{1/2},
\end{align}
where $\rho_{\rm cr}/s_{\rm now}\simeq 3.6\,h^2\times 10^{-9}$ is 
the ratio of critical density to the current entropy density of Universe, and $h$ is the 
dimensionless Hubble parameter. 
Here, we suppose that the annihilation cross section of LSPs 
is of ${\cal O}(10^{-7}-10^{-8}\,{\rm GeV}^{-2})$, which is consistent with 
those of wino-like neutralino~\cite{Moroi:1999zb},
\begin{align}
\langle \sigma_{\rm ann}v\rangle 
\simeq \frac{g_2^4}{2\pi}\frac{m_{\rm LSP}^2}{(2m_{\rm LSP}^2-m_W^2)^2}
\left( 1-\frac{m_W^2}{m_{\rm LSP}^2}\right)^{3/2},
\end{align}
with $m_W$ being the $W$-boson mass, 
and Higgsino-like neutralino into the $W$-boson pair~\cite{Olive:1989jg},
\begin{align}
\langle \sigma_{\rm ann}v\rangle 
\simeq \frac{g_2^4}{32\pi}\frac{m_{\rm LSP}^2}{(2m_{\rm LSP}^2-m_W^2)^2}
\left( 1-\frac{m_W^2}{m_{\rm LSP}^2}\right)^{3/2},
\end{align}
in the absence of co-annihilation effect.  
Therefore, the dark matter abundance in the low-scale SUSY-breaking scenario is 
overabundant to the Planck result, $0.1175 \le \Omega_{\rm LSP} h^2 \le 0.1219$~\cite{Ade:2013zuv,Ade:2015lrj}, 
unless the branching ratio is smaller than ${\cal O}(10^{-4})$. 

On the other hand, in the high-scale SUSY-breaking scenario, 
the second term in Eq.~(\ref{eq:dmyield})  dominates the LSP yield. 
The dark matter abundance is then given by 
\begin{align}
\Omega_{\rm LSP}h^2&\simeq 
m_{\rm LSP} 
\sqrt{\frac{45}{8\pi^2g_\ast (T_{3/2})}} \frac{1}{M_{\rm Pl}T_{3/2}\langle \sigma_{\rm ann}v\rangle}
\frac{s_{\rm now}}{\rho_{\rm cr}}
\nonumber\\
&\simeq 65 \left(\frac{80}{g_\ast (T_{3/2}) }\right)^{1/2}
\left(\frac{10^{6}\,{\rm GeV}}{m_{3/2}}\right)^{3/2}
\left( \frac{m_{\rm LSP}}{1\,{\rm TeV}}\right)^3
\left(\frac{10^{-3}\,{\rm GeV}^{-2}}{m_{\rm LSP}^2\langle \sigma_{\rm ann}v\rangle}\right),
\end{align} 
where the dark matter is assumed to be the wino-like or Higgsino-like neutralinos. 
Since, in the gravity-mediated SUSY-breaking scenario, the mass of LSP increases as a 
consequence of large gravitino mass, 
the overabundance of dark matter is common feature in the moduli-dominated Universe. 

So far, there are several studies to dilute the gravitino abundance via the thermal 
inflation~\cite{Lyth:1995ka}, Q-ball~\cite{Fujii:2001xp} and 
unstable domain-wall~\cite{Hattori:2015xla} or the introduction of the axion 
sector~\cite{Nakamura:2008ey}. The modulus oscillation may be suppressed 
by the adiabatic oscillation~\cite{Linde:1996cx,Nakayama:2011wqa}. 
In the next sections~\ref{sec:SUGRA} and~\ref{sec:typeII}, 
we show the new dilution mechanism by introducing an extra chiral multiplet.

\section{The dilution mechanism in $4$D ${\cal N}=1$ SUGRA}
\label{sec:SUGRA}
Here, we propose a new dilution mechanism based on 
the $4$D ${\cal N}=1$ SUGRA. To reduce the gravitino abundance 
by the modulus decay, we introduce  another chiral multiplet 
$\Phi$ with the following ansatz of (real) K\"ahler potential and 
superpotential\footnote{Although $\Phi$ is identified as the complex structure modulus field 
in the next section, the extension to other models are straightforward.},
\begin{align}
K&=-3\ln (-i(T-\bar{T})) +K(\Phi, \bar{\Phi}),
\nonumber\\
W&=w_0 +Ae^{iaT}.
\end{align}
In terms of the above K\"ahler potential and 
superpotential, the {\it F}-term scalar potential on the basis of $4$D 
${\cal N}=1$ SUGRA is described by
\begin{align}
V=e^K\left( K^{I\bar{J}} D_IW D_{\bar{J}} \bar{W} -3|W|^2\right),
\end{align}
where $D_IW =W_I +K_I W$, $K_I=\partial K/\partial \Phi^I$, $W_I=\partial W/\partial \Phi^I$ 
with $\Phi^I=T, \Phi$, and $K^{I\bar{J}}$ is the inverse of the 
K\"{a}hler metric $K_{I\bar{J}}$. 
Since there is no kinetic mixing between $T$ and $\Phi$, 
they are independently stabilized at the supersymmetric AdS minimum 
satisfying
\begin{align}
D_TW&=0,
\nonumber\\
D_\Phi W&=K_{\Phi}W=0.
\label{eq:Phi}
\end{align}

To uplift the AdS vacuum to the dS vacuum with tiny cosmological constant, 
we require the certain uplifting scenario as discussed in the next section. 
If we assume that the uplifting sector does not depend on the added chiral superfield $\Phi$, 
$\Phi$ still stays at the supersymmetric minimum given by Eq.~(\ref{eq:Phi}). 
In this supersymmetric minimum, the mass squared of modulus field is found 
in the limit of $a\langle T_{\cal I}\rangle \gg 1$,
\begin{align}
m_T^2 &\simeq \frac{\partial_T\partial_{\bar{T}} V}{K_{T\bar{T}}} 
\simeq 4(a\langle T_{\cal I}\rangle)^2 m_{3/2}^2,
\label{eq:Tmass}
\end{align}
whereas the mass matrix of $\Phi=\Phi_{\cal R} +i \Phi_{\cal I}$ is given by 
\begin{align}
m_\Phi^2 &=(2K_{\Phi \bar{\Phi}})^{-1}
\begin{pmatrix}
\partial_{\Phi_{\cal R}} \partial_{\Phi_{\cal R}} V & \partial_{\Phi_{\cal R}} \partial_{\Phi_{\cal I}} V\\
\partial_{\Phi_{\cal R}} \partial_{\Phi_{\cal I}} V & \partial_{\Phi_{\cal I}} \partial_{\Phi_{\cal I}} V
\end{pmatrix}
\nonumber\\
 &=\frac{m_{3/2}^2}{4}(K^{\Phi \bar{\Phi}})^{2}
\begin{pmatrix}
4|K_{\Phi\Phi_{\cal R}}|^2  & \left(K_{\Phi_{\cal R}\Phi_{\cal R}}+K_{\Phi_{\cal I}\Phi_{\cal I}}\right)K_{\Phi_{\cal R}\Phi_{\cal I}}\\
\left(K_{\Phi_{\cal R}\Phi_{\cal R}}+K_{\Phi_{\cal I}\Phi_{\cal I}}\right)K_{\Phi_{\cal R}\Phi_{\cal I}} 
& 4|K_{\Phi\Phi_{\cal I}}|^2
\end{pmatrix}
.
\end{align}
When the kinetic mixing between the real and imaginary parts of $\Phi$ is absent 
at the minimum, 
i.e., $\langle K_{\Phi_{\cal R}\Phi_{\cal I}}\rangle =0$, 
we always obtain the positive mass squared of $\Phi$. 
Furthermore, with ${\cal O}(1)$ value of K\"ahler metric, $K_{\Phi\bar{\Phi}}={\cal O}(1)$, 
the modulus mass is typically larger than that of $\Phi$,
\begin{align}
m_{\Phi_{\cal R}}^2 \simeq m_{\Phi_{\cal I}}^2 \simeq m_{3/2}^2.
\end{align}
We remark that 
the above positive mass squared of $\Phi$ cannot be realized at the AdS 
minimum without the uplifting sector.(See for Ref.~\cite{Conlon:2006tq} 
in the situation that $\Phi$ corresponds to the no-scale modulus.)

Let us take a closer look at the dynamics of light added field $\Phi$. 
As we mentioned above, the modulus field $T$ gravitationally couples to the matter fields 
and then the minimum of modulus field during the inflation, $T_1$, is generically 
different from the minimum given in Eq.~(\ref{eq:Phi}), 
which are represented by $T_0$ and $\Phi_0$ in what follows. 
Moreover, the gravitational interaction between $T$ and inflaton sector 
induces the Planckian distance between $T_1$ and $T_0$. 
When the Hubble scale $H$ is comparable to the mass scale of $T$, $m_{T}$, 
modulus field rolls down to the true minimum and dominates the energy density of 
the Universe as mentioned in Sec.~\ref{sec:review}. 
In a  way similar to the modulus field, 
we consider that the added field $\Phi$ only gravitationally couples to the matter fields 
on the same footing. 

Such a situation is captured by the following simplified scalar 
potential\footnote{We do not consider the adiabatic suppression scenario where 
the coefficient of Hubble parameter is much larger than unity~\cite{Linde:1996cx,Nakayama:2011wqa}.},
\begin{align}
V_{\rm osc}=\frac{H^2(t)}{2} |T -T_1|^2 +\frac{m_{T}^2}{2} |T-T_0|^2 +
\frac{H^2(t)}{2} |\Phi -\Phi_1|^2 +\frac{m_{\Phi}^2}{2} |\Phi -\Phi_0|^2,
\end{align}
where $T_1$ and $\Phi_1$ denote the minimum induced by the Hubble parameter $H(t)$. 
Here, we denote the canonically normalized fields of $T$ and $\Phi$ 
as the same notation $T$ and $\Phi$, respectively. 
Since, in our model, the added field $\Phi$ is lighter than the modulus $T$, 
$\Phi$ oscillates subsequent to the oscillation of $T$, i.e., $t_{\rm osc}^\Phi > t_{\rm osc}^T$, 
where the oscillating time of scalar fields $\Phi$ and $T$ are 
defined as $t_{\rm osc}^\Phi$ and $t_{\rm osc}^T$, respectively. 
In the regime $t_{\rm osc}^T < t <t_{\rm osc}^\Phi$, the equation of motion of $T$ reads,
\begin{align}
&\frac{d^2T}{dt^2} +3H \frac{dT}{dt} +m_T^2 (T-T_0)=0,\nonumber\\
&3H^2 \simeq \frac{1}{2}\left|\frac{dT}{dt}\right|^2 +\frac{m_{T}^2}{2} |T-T_0|^2.
\label{eq:eqmT}
\end{align}
When we redefine the modulus field as $\tilde{T}=a^{3/2}(T-T_0)$, 
Eq.~(\ref{eq:eqmT}) is rewritten as
\begin{align}
&\frac{d^2\tilde{T}}{dt^2} +\left(m_T^2-\frac{3}{2}\frac{dH}{dt} -\frac{9}{4}H^2\right) \tilde{T}=0,
\end{align}
which can be solved under $m_T >\{H, |d H/d t| \}$,
\begin{align}
&\tilde{T}(t)=\tilde {T}_0\,{\rm sin}(m_T t),
\end{align}
with $\tilde{T}_0$ being constant. 
Furthermore, the virial theorem, 
$\frac{1}{2}\left\langle \frac{dT}{dt}\right\rangle^2 =\frac{1}{2}\left\langle m^2T^2\right\rangle 
=a^{-3}m^2\tilde{T}_0^2/4$, enables us to solve the time evolution of modulus field $T(t)$,
\begin{align}
&T(t)\simeq \sqrt{\frac{8}{3}}\frac{M_{\rm Pl}}{m_T t}\,{\rm sin}(m_T t),
\end{align}
from which the initial displacement $\Delta T=T_1-T_0$ is assumed to be of ${\cal O}(M_{\rm Pl})$ 
at the time $t_{\rm osc}^T\simeq 1/m_T$. 
The light field $\Phi$ also oscillates at the time $t_{\rm osc}^\Phi \simeq 1/m_\Phi$ in 
a similar fashion. 
When we assume that the initial displacement of $\Phi$ as $\Delta \Phi={\cal O}(M_{\rm Pl})$, 
the oscillating energy densities of $\Phi$ and $T$ at the time $t_{\rm osc}^\Phi \simeq 1/m_\Phi$ 
are almost the same as each other,
\begin{align}
\frac{1}{2}m_T^2T(t_{\rm osc}^\Phi)^2\simeq \frac{1}{2}m_\Phi^2 (\Delta \Phi)^2. 
\end{align}
It implies that two scalar fields dominate the energy densities of the Universe at the time $t_{\rm osc}^\Phi$. 

Next, let us move onto the reheating process of $T$ and $\Phi$. 
As shown in Eq.~(\ref{eq:Ttot}), the modulus field dominantly decays into 
the gauge boson pairs. 
By contrast, although we do not determine the couplings between $\Phi$ and standard 
model sector, for the time being, we assume that the decay temperature of $\Phi$ is 
much smaller than that of modulus field $T$.(We will provide the 
detailed setup in the next section.) 
To simplify our analysis, we further assume that decay temperatures of real and imaginary parts of 
$\Phi$ are the same. 
Therefore, the lightest field, $\Phi$, dominates the energy 
density of the Universe at the time later than the decay 
time of $T$, $t_{\rm dec}^T\simeq 1/\Gamma_{\rm tot}^{T}$.\footnote{Note that the decay temperatures of real and imaginary parts of 
modulus are also the same as shown in Eq.~(\ref{eq:Ttot}).} 
Keeping mind that $\Phi$ has the vanishing  {\it F}-term, 
we find that this field does not decay into 
the gravitino(s). 
At the decay time of $\Phi$, $t_{\rm dec}^{\Phi}\simeq 1/\Gamma_{\rm tot}^{\Phi}$ with $\Gamma_{\rm tot}^{\Phi}$ 
being the total decay width of $\Phi$, the produced entropy dilutes 
the gravitino (and LSP) yield given through the decay of $T$ at the time $t_{\rm dec}^T$. 
The dilution factor defined by the ratio between the entropy density of $\Phi$ ($s_{\Phi}$), 
and that of $T$ ($s_T$), at the time $t_{\rm dec}^{\Phi}$ is given by
\begin{align}
\Delta_{S} &\simeq \frac{s_{\Phi}(t_{\rm dec}^{\Phi})}{s_{T}(t_{\rm dec}^{\Phi})} 
\simeq \left(\frac{\rho_{\Phi}(t_{\rm dec}^{\Phi})}{\rho_{T}(t_{\rm dec}^{\Phi})}\right)^{3/4}
\simeq \left(\frac{\rho_{\Phi}(t_{\rm dec}^{\Phi})}{\rho_{\Phi}(t_{\rm dec}^{T})}
\frac{\rho_{T}(t_{\rm dec}^{T})}{\rho_{T}(t_{\rm dec}^{\Phi})}
\frac{\rho_{\Phi}(t_{\rm dec}^{T})}{\rho_{T}(t_{\rm dec}^{T})}
\right)^{3/4}
\nonumber\\
& \simeq \Bigl[ \left(\frac{a(t_{\rm dec}^{T})}{a(t_{\rm dec}^{\Phi})}\right)^3
\left(\frac{a(t_{\rm dec}^{\Phi})}{a(t_{\rm dec}^{T})}\right)^4
\Bigl]^{3/4}
\simeq \left(\frac{a(t_{\rm dec}^{\Phi})}{a(t_{\rm dec}^{T})}\right)^{3/4}
\simeq \left(\frac{t_{\rm dec}^{\Phi} }{t_{\rm dec}^T}\right)^{1/2}
\nonumber\\
&
\simeq \left(\frac{\Gamma_{\rm tot}^T}{\Gamma_{\rm tot}^{\Phi}}\right)^{1/2},
\end{align}
which is much larger than unity under our assumption. 

In this way, even if a lot of gravitino is generated at the modulus decay, 
huge entropy injection dilutes the gravitino and LSP abundance. 
Although we assume that $\Phi$ gravitationally couples to the matter fields 
and decays into them after the decay of modulus field until now, 
we expect that proposed dilution 
mechanism is applied in the general class of models. 
If there is a sizable kinetic mixing between $T$ and $\Phi$, 
the lightest scalar field which is the linear combination of $T$ and $\Phi$ 
would decays into the gravitino(s) after diagonalizing them. 
Note that, supersymmetric stabilization condition, $D_{\Phi}W=0$, 
is also discussed in the gravitino production from the inflaton decay~\cite{Kawasaki:2006gs}. 
In the next section, we demonstrate the above analysis within the framework of 
type II string theory.

\section{The dilution mechanism in effective action of type II string theory}
\label{sec:typeII}
To make the analysis concrete, we show the dilution mechanism based on the effective action of 
type II string theory, e.g., type IIB string theory. 
As pointed out in Ref.~\cite{Giddings:2001yu}, we have often assumed that all the complex structure moduli and axion-dilaton are stabilized at the compactification scale by the three-form fluxes. 
However, the above statement depends on the ansatz of three-form fluxes. 
In this section, we assume that one of the complex structure moduli  remains massless 
at the perturbative level in the flux compactification. 
In this case, this massless moduli ($U$) appears through the non-perturbative effects in 
the prepotential (See for more details, e.g., Ref.~\cite{Hori:2000kt}.),
\begin{align}
F=F_{\rm pert} -\frac{n}{(2\pi i)^3} e^{2\pi iU},
\end{align}
where $F_{\rm pert}$ is the perturbative prepotential, and 
$n$ is the instanton number associated with the relevant cycle. 
Below the mass scales of other stabilized  complex structure moduli and axion-dilaton, 
the K\"ahler potential in Eq.~(\ref{eq:Kahler}) reduces to be
\begin{align}
K&=-3\ln (-i(T-\bar{T})) \nonumber\\
&-\ln \Bigl[ U_{\cal I}\beta +\gamma +\frac{n}{2\pi^2}U_{\cal I} e^{-2\pi U_{\cal I}}{\rm cos}(2\pi U_{\cal R}) 
+\frac{n}{2\pi^3}e^{-2\pi U_{\cal I}}{\rm cos}(2\pi U_{\cal R})  \Bigl],
\label{eq:Kinst}
\end{align}
where $\beta$ and $\gamma$ are the positive real constant determined by the 
vacuum expectation values of other stabilized moduli fields. 
Here, we consider the vanishing self-intersection number for $U=U_{\cal R}+iU_{\cal I}$, for simplicity. 
It is straightforward to extend our following discussion to the case of nonvanishing self-intersection 
number of $U$. 

Let us further assume that the superpotential does not depend on the remaining complex structure 
modulus $U$. 
This situation is ensured by choosing the $U$-independent three-form fluxes in Eq.~(\ref{eq:fluxW}) 
and assuming that the threshold correction to the gauge coupling in the hidden sector is independent of $U$~\cite{Dixon:1990pc,Lust:2003ky}. 
In terms of the effective superpotential described in Eq.~(\ref{eq:KW}) and 
K\"ahler potential in Eq.~(\ref{eq:Kinst}), 
the {\it F}-term scalar potential on the basis of $4$D ${\cal N}=1$ supergravity 
is given by
\begin{align}
V=e^K\left( K^{I\bar{J}} D_IW D_{\bar{J}} \bar{W} -3|W|^2\right),
\end{align}
where $D_IW =W_I +K_I W$, $K_I=\partial K/\partial \Phi^I$, $W_I=\partial W/\partial \Phi^I$ 
with $\Phi^I=T, U$, and $K^{I\bar{J}}$ is the inverse of the 
K\"{a}hler metric $K_{I\bar{J}}$. 

Although the stabilization condition of K\"ahler modulus is the same as that of KKLT moduli 
stabilization, $D_TW=0$, the complex structure modulus can be stabilized at the minimum 
satisfying
\begin{align}
K_{U_{\cal I}}&= -\frac{\beta -\frac{n}{2\pi^2}e^{-2\pi U_{\cal I}}{\rm cos}(2\pi U_{\cal R}) 
-\frac{n}{\pi}U_{\cal I}e^{-2\pi U_{\cal I}}{\rm cos}(2\pi U_{\cal R})}{U_{\cal I}\beta +\gamma +\frac{n}{2\pi^2}U_{\cal I} e^{-2\pi U_{\cal I}}{\rm cos}(2\pi U_{\cal R}) 
+\frac{n}{2\pi^3}e^{-2\pi U_{\cal I}}{\rm cos}(2\pi U_{\cal R})}=0,
\nonumber\\
K_{U_{\cal R}}&=\frac{\frac{n}{\pi}U_{\cal I} e^{-2\pi U_{\cal I}}{\rm sin}(2\pi U_{\cal R}) 
+\frac{n}{\pi^2}e^{-2\pi U_{\cal I}}{\rm sin}(2\pi U_{\cal R})}
{U_{\cal I}\beta +\gamma +\frac{n}{2\pi^2}U_{\cal I} e^{-2\pi U_{\cal I}}{\rm cos}(2\pi U_{\cal R}) 
+\frac{n}{2\pi^3}e^{-2\pi U_{\cal I}}{\rm cos}(2\pi U_{\cal R}) }=0,
\end{align}
which are solved as 
\begin{align}
\langle U_{\cal R}\rangle&=0,\nonumber\\
\beta -\frac{n}{2\pi^2}e^{-2\pi \langle U_{\cal I}\rangle}
-\frac{n}{\pi}\langle U_{\cal I}\rangle e^{-2\pi \langle U_{\cal I}\rangle}&=0.
\label{eq:URI}
\end{align}
With the parameters $\beta=1$ and $n=1000$, 
the vacuum expectation value of $U_{\cal I}$ becomes
\begin{align}
\langle U_{\cal I}\rangle\simeq 0.93,
\label{eq:UIvev}
\end{align}
in the string unit. 
Such numerical values of parameters and moduli vacuum expectation values are ensured as follows. 
As shown in Refs.~\cite{Hosono:1993qy, Hosono:1994ax}, the large instanton number often appears in 
a certain CY manifold. 
It thus allows us to treat the small vacuum expectation values of complex structure modulus 
compared with the string length, since we treat the quantum-corrected prepotential. 
Note that K\"ahler and complex structure moduli are independently stabilized 
at the supersymmetric minimum, $D_TW=D_UW=0$. 

However, the obtained supersymmetric minimum gives rise to 
a negative energy density of the scalar potential, i.e., AdS vacuum. 
In order to uplift the vacuum to dS vacuum, 
there are several uplifting scenario such as {\it F}-term 
uplifting~\cite{Dudas:2006gr,Abe:2006xp,Kallosh:2006dv,Abe:2007yb},
and explicit SUSY-breaking as an existence of anti D-brane~\cite{Kachru:2003aw}. 
Let us analyze the moduli masses and SUSY-breaking sector for each individual case.

\subsection{{\it F}-term uplifting}
\label{subsec:Fterm}
The SUSY-breaking sector living on hidden D-brane enables us to uplift the AdS vacuum 
to the dS vacuum with tiny cosmological constant~\cite{Dudas:2006gr,Abe:2006xp,Kallosh:2006dv,Abe:2007yb}. 
In this section, we study three types of moduli stabilization with F-term uplifting scenario 
by the SUSY breaking such as  the Intriligator-Seiberg-Shih scenario \cite{Intriligator:2006dd}.

\subsubsection{ Model 1}
\label{subsubsec:ISSKKLT}
First of all, we consider the following K\"ahler and superpotential on hidden sector 
in addition to the moduli K\"ahler potential in Eq.~(\ref{eq:Kinst}) and superpotential in Eq.~(\ref{eq:KW}),
\begin{align}
\Delta K &=Z(T-\bar{T}) |X|^2 -Z^{(1)}(T-\bar{T})\frac{|X|^4}{\Lambda^2},
\nonumber\\
\Delta W&=\mu X,
\label{eq:X}
\end{align}
where $\mu$ is the real constant determined by the heavy moduli fields, 
and the four-point coupling of the SUSY-breaking multiplet $X$ in the 
K\"ahler potential appears after integrating out the heavy mode with mass $\Lambda$.  
We now assume that the K\"ahler metric of SUSY-breaking multiplet $X$ is 
independent of $U$, otherwise $U$  obtains the {\it F}-term. 
We come back to the case of $U$-dependent uplifting scenario in Appendix~\ref{app:uplift}. 
From the K\"ahler potential~(\ref{eq:Kinst}) and superpotential~(\ref{eq:KW}) including the 
SUSY-breaking sector~(\ref{eq:X}) at the supersymmetric minimum of moduli 
fields, $D_TW=D_UW=0$, 
the extremal condition of SUSY-breaking field $X$, 
\begin{align}
\frac{\partial V}{\partial \bar{X}}&\simeq \partial_{\bar X} \Biggl[ e^K\left( 
K^{X\bar{X}} |D_X W|^2-3|W|^2\right)\Biggl] 
\simeq e^K \left( \partial_{\bar{X}} (K^{X\bar{X}}) \mu^2 -2\mu w\right)
\simeq  0,
\end{align}
leads to the minimum of $X$,
\begin{align}
\langle X\rangle \simeq \frac{Z(T-\bar{T})^2w}{2Z^{(1)}(T-\bar{T})\mu}\Lambda^2,
\label{eq:Xref}
\end{align}
which is much smaller than unity under $\Lambda \ll 1$. 
At this minimum, the tiny cosmological constant can be realized by properly choosing $\mu$ and $w$ as
\begin{align}
&\langle V\rangle \simeq e^K\Biggl[ K^{X\bar{X}} |D_X W|^2 -3|W|^2\Biggl] 
\simeq e^K\left( \frac{\mu^2}{Z(T-\bar{T})} -3w^2 \right)\simeq 0.
\end{align}

However, the inclusion of SUSY-breaking sector violates the extremal conditions for the moduli fields. 
To find the true minimum of moduli fields ($T$, $U$), 
we evaluate the deviations from the supersymmetric minimum, $\delta T=T-T_{\rm ref}$ and 
$\delta U=U-U_{\rm ref}$ where $T_{\rm ref}$ and $U_{\rm ref}$ are the reference points 
satisfying the supersymmetric conditions, $D_TW|_{\rm ref}=D_UW|_{\rm ref}=0$. 
The true minimum of SUSY-breaking field $X$ is also found by evaluating 
the deviation from the reference point in Eq.~(\ref{eq:Xref}), $\delta X=X-\langle X\rangle$. 
First of all, let us take a closer look at $\delta U$. 
As far as the SUSY-breaking field does not couple to $U$, 
the extremal condition of $U$, $\partial V/\partial U=0$, is always 
satisfied under the supersymmetric condition, $K_U=0$, i.e., $\delta U=0$. 

Next, we evaluate the variations, $\delta T$ and $\delta X$, in terms of 
\begin{align}
&D_TW =W_{TT}|_{\rm ref} \delta T +(K_T W_T)|_{\rm ref}\delta T +(K_T W_X)|_{\rm ref}\delta X
+( K_{T\bar{T}} W)|_{\rm ref} 
\left( \delta \bar{T}-\delta T\right),
\nonumber\\
&D_XW\simeq W_X|_{\rm ref} +(K_X W_T)|_{\rm ref}\delta T
+( K_{X\bar{X}} W)|_{\rm ref} 
\left( \delta \bar{X}-\delta X\right) 
,
\end{align}
where we take the limit $\langle X\rangle \ll 1$.
From the scalar potential at quadratic order of $\delta T$ and $\delta X$, 
\begin{align}
V =&V|_{\rm ref} +V_I|_{\rm ref} \delta\phi^I +V_{\bar{I}}|_{\rm ref} \bar{\delta\phi^I}
\nonumber\\
&+\frac{1}{2}V_{IJ}|_{\rm ref} \delta\phi^I \delta\phi^J +V_{I\bar{J}}|_{\rm ref} \delta\phi^I \bar{\delta\phi^J} +\frac{1}{2}V_{\bar{I}\bar{J}}|_{\rm ref} \bar{\delta\phi^I} \bar{\delta\phi^J}, 
\end{align}
with $V_I|_{\rm ref}=\partial_I V|_{\rm ref}$ and $V_{IJ}|_{\rm ref}=\partial_I\partial_J V|_{\rm ref}$ being 
the first and second derivatives with respect to the fields, $\phi^I =T,X$, at their reference points, 
we obtain the variations of $T$ and $X$ 
\begin{align}
\delta T&\simeq \frac{9|W|^2}{2T_{\cal I}K^{T\bar{T}}|W_{TT}|^2}\simeq 
\frac{3i}{2T_{\cal I} a^2},
\nonumber\\
\delta X&\simeq \frac{\sqrt{3}Z(T-\bar{T})^{3/2}}{6Z^{(1)}(T-\bar{T})}\Lambda^2,
\end{align}
in which our analysis is justified due to the following equality:
\begin{align}
\Bigl|V_I|_{\rm ref} \delta\phi^I +V_{\bar{I}}|_{\rm ref} \bar{\delta\phi^I} \Bigl| \gg 
\Bigl|\frac{1}{2}V_{IJ}|_{\rm ref} \delta\phi^I \delta\phi^J +V_{I\bar{J}}|_{\rm ref} \delta\phi^I \bar{\delta\phi^J} +\frac{1}{2}V_{\bar{I}\bar{J}}|_{\rm ref} \bar{\delta\phi^I} \bar{\delta\phi^J}\Bigl|.
\end{align}
Since there is no kinetic mixing between $U$ and $T$, 
the mass squared of canonically normalized moduli fields are evaluated at the 
obtained minimum, $T=T_{\rm ref} +\delta T$, $U=U_{\rm ref}$, 
and $X=X_{\rm ref} +\delta X$, 
\begin{align}
m_{U_{\cal R}}^2&=\frac{V_{U_{\cal R}U_{\cal R}}}{2K_{U\bar{U} }}=\frac{1}{4}\left( \frac{K_{U_{\cal R}U_{\cal R}}}{ K_{U\bar{U}}}\right)^2 
m_{3/2}^2
=(2\pi)^2 \left(U_{\cal I} +\frac{1}{\pi}\right)^2m_{3/2}^2,
\nonumber\\
m_{U_{\cal I}}^2&=\frac{V_{U_{\cal I}U_{\cal I}}}{2K_{U\bar{U} }}=\frac{1}{4}\left( \frac{K_{U_{\cal I}U_{\cal I}}}{ K_{U\bar{U}}}\right)^2 
m_{3/2}^2
=(2\pi)^2 U_{\cal I}^2m_{3/2}^2,
\nonumber\\
m_{T_{\cal R}}^2 &=\frac{V_{T_{\cal R}T_{\cal R}}}{2K_{T\bar{T}}}=
\frac{3a^2}{2K_{T\bar{T}}}m_{3/2}^2
\simeq (2a T_{\cal I})^2 m_{3/2}^2,
\nonumber\\
m_{T_{\cal I}}^2 &\simeq m_{T_{\cal R}}^2,
\label{eq:TUmass}
\end{align}
whereas that of SUSY-breaking field is given by 
$m_{X_{\cal R}}^2 \simeq m_{X_{\cal I}}^2 \simeq \frac{6Z^{(1)}(T-\bar{T})}{Z(T-\bar{T})^2\Lambda^2}m_{3/2}^2$, which can be taken larger than the moduli masses. 
Therefore, moduli decay into the dynamical SUSY-breaking sector is neglected. 
With $U_{\cal I} <1$ and $a T_{\cal I} \sim 4\pi^2$, 
the mass squared of complex structure modulus is typically smaller than that of K\"ahler modulus. 
Furthermore, the corresponding {\it F}-terms are given by
\begin{align}
\left\langle\frac{F^T}{T-\bar{T}}\right\rangle &
\simeq \frac{3}{2aT_{\cal I}}m_{3/2} 
\simeq 3\frac{m_{3/2}^2}{m_{T_{\cal I}}},
\nonumber\\
\langle F^U\rangle &=0,
\nonumber\\
\left\langle F^X\right\rangle &\simeq -\sqrt{\frac{3}{Z(T-\bar{T})}}m_{3/2}.
\end{align}
As a result, the complex structure modulus lighter than K\"ahler modulus does not violate the SUSY 
at the vacuum. 
When the SUSY-breaking field has $U$-dependent K\"ahler metric in Eq.~(\ref{eq:X}), 
the complex structure modulus induces the SUSY-breaking as studied in Appendix~\ref{app:uplift}.

\subsubsection{ Model 2} 
\label{subsubsec:ISSRT}
Next, we comment on  another K\"ahler 
moduli stabilization, in contrast to the KKLT-type. 
The detail of following procedure is the same as previous 
one in Sec.~\ref{subsubsec:ISSKKLT}\footnote{It is summarized in e.g., 
Refs.~\cite{Abe:2006xp,Abe:2007yb}.}. 
When the K\"ahler modulus is stabilized at the racetrack minimum 
in the superpontential, 
\begin{align}
W= w_0 +Be^{ibT} -Ce^{icT},
\label{eq:rtrack}
\end{align}
where two non-perturbative effects are generated associated with 
the cycle $T$ through the hidden extra D$7$-branes with $b \sim c \sim 4\pi^2$, 
the supersymmetric minimum of modulus field is found as
\begin{align}
T_{\rm ref}\sim \frac{1}{b-c}\ln \left(\frac{bB}{cC}\right).
\label{eq:RT}
\end{align}
By combing the SUSY-breaking sector in Eq.~(\ref{eq:X}) and racetrack scenario, 
we find that the deviations from the reference points given in Eqs.~(\ref{eq:Xref}) 
and~(\ref{eq:RT}) are 
\begin{align}
\delta T&\simeq i\frac{9|W|^2}{2T_{\cal I}K^{T\bar{T}}|W_{TT}|^2},
\nonumber\\
\delta X&\simeq \frac{\sqrt{3}Z(T-\bar{T})^{3/2}}{6Z^{(1)}(T-\bar{T})}\Lambda^2,
\end{align}
at which the mass squared of K\"ahler modulus and the corresponding {\it F}-term 
become 
\begin{align}
m_{T_{\cal I}}^2 &\simeq m_{T_{\cal R}}^2 \simeq 
\frac{16Z(T-\bar{T})}{3}(bc)^2 T_{\cal I}^4m_{3/2}^2 
\biggl[\left(1-\frac{b}{c}\right) \frac{Be^{-bT_{\cal I}}}{\mu}\biggl]^2,
\nonumber\\
\left\langle\frac{F^T}{T-\bar{T}}\right\rangle &\simeq 
\frac{3\left(bBe^{-bT_{\cal I}}-cCe^{-cT_{\cal I}}\right)}{2T_{\cal I}
\left(-b^2 Be^{-bT_{\cal I}}+c^2 Ce^{-cT_{\cal I}}\right)}m_{3/2}
\simeq 3\frac{m_{3/2}^2}{m_{T_{\cal I}}}.
\end{align}
The mass squared and {\it F}-term of complex structure modulus and SUSY-breaking field are 
the same as in Sec.~\ref{subsubsec:ISSKKLT}. 
Thus, in contrast to the single non-perturbative effect for $T$, 
the K\"ahler modulus is much heavier than the light complex structure modulus. 
According to it, the {\it F}-term of  K\"ahler modulus is more suppressed than 
the case of single non-perturbative effect.

\subsubsection{ Model 3}
\label{subsubsec:SISS}
Finally, we discuss another {\it F}-term uplifting scenario combined with the 
racetrack superpotential in Eq.~(\ref{eq:rtrack}). 
The K\"ahler potential and superpotential of hidden sector are described by
\begin{align}
\Delta K &=Z(T-\bar{T}) |X|^2 -Z^{(1)}(T-\bar{T})\frac{|X|^4}{\Lambda^2},
\nonumber\\
\Delta W&=De^{idT} X,
\label{eq:X2}
\end{align}
where the above superpotential can be generated by the gaugino condensation 
and D-brane instanton effects. 
In a similar way to the discussion in Secs.~\ref{subsubsec:ISSKKLT} and~\ref{subsubsec:ISSRT}, 
the reference points of moduli fields and SUSY-breaking fields are the same as one obtained above. 
However, the superpotential~(\ref{eq:X2}) leads to the following deviations from the reference points,
\begin{align}
\delta T&\simeq i\frac{3d|W|^2}{K^{T\bar{T}}|W_{TT}|^2} 
\simeq i\frac{9d|W|^2}{4T_{\cal I}^2 (-b^2Be^{-bT_I}+c^2Ce^{-cT_I})^2},
\nonumber\\
\delta X&\simeq \frac{\sqrt{3}Z(T-\bar{T})^{3/2}}{6Z^{(1)}(T-\bar{T})}\Lambda^2 
\left[ 1+\frac{\sqrt{3}d^2W}{2(-b^2Be^{-bT_I}+c^2Ce^{-cT_I})}\right],
\end{align}
at which the mass squared of K\"ahler modulus and the corresponding {\it F}-term 
become 
\begin{align}
m_{T_{\cal I}}^2 &\simeq m_{T_{\cal R}}^2 \simeq 
\frac{16Z(T-\bar{T})}{3}(bc)^2 T_{\cal I}^4m_{3/2}^2 
\biggl[ \frac{B}{D}\left(1-\frac{b}{c}\right) e^{-(b-d)T_{\cal I}}\biggl]^2,
\nonumber\\
\left\langle\frac{F^T}{T-\bar{T}}\right\rangle &\simeq 
\frac{3d}{2T_{\cal I}}\frac{m_{3/2}^2}{m_{T_{\cal I}}}.
\end{align}
The mass squared and {\it F}-term of complex structure modulus and SUSY-breaking field are 
the same as in Sec.~\ref{subsubsec:ISSKKLT}, whereas 
the mass and {\it F}-term of K\"ahler modulus are different from the previous setups in 
Secs.~\ref{subsubsec:ISSRT} and~\ref{subsubsec:SISS}. 
With the parameter $d\simeq 4\pi^2$, 
it turns out that only the {\it F}-term is taken to be larger than that in 
Sec.~\ref{subsubsec:ISSRT}.

\subsection{Uplifting with anti D-brane}
\label{subsec:uplift}
In this section, we show  another uplifting scenario 
as an existence of anti D-brane~\cite{Kachru:2003aw} 
located at a certain warped throat. 
Tiny cosmological constant is realized from the positive vacuum energy 
originating from the anti D-branes, although these anti D-branes induce 
the explicit SUSY-breaking in contrast to the spontaneous SUSY-breaking 
in Sec.~\ref{subsec:Fterm}. 
As proposed in Refs.~\cite{Choi:2004sx,Choi:2005ge}, 
the uplifting term is formulated in the $4$D effective supergravity,
\begin{align}
{\cal L}_{\rm up}=-\int d^4\theta |C|^4 \theta^2\bar{\theta}^2{\cal P}, 
\end{align} 
where ${\cal P}$ denotes the unknown function of moduli fields. 
Then, we obtain the uplifting scalar potential is given by
\begin{align}
V_{\rm up} =e^{2K/3}{\cal P}.  
\end{align} 
When the ${\cal P}$ is independent of $U$, the complex 
structure modulus $U$ cannot be deviated from the supersymmetric 
minimum, $K_U=0$. However, the mass squared of $U$ receives the 
contribution from the uplifting sector,
\begin{align}
m_{U_{\cal R}}^2&=(2\pi)^2 \left( U_{\cal I} +\frac{1}{\pi}\right)^2m_{3/2}^2 
+4\pi \left( U_{\cal I} +\frac{1}{\pi}\right)m_{3/2}^2,
\nonumber\\
m_{U_{\cal I}}^2&=(2\pi)^2 U_{\cal I}^2m_{3/2}^2 +4\pi U_{\cal I} m_{3/2}^2.
\end{align}
Thus, the SUSY is only broken by the 
K\"ahler moduli. 
Along the same step outlined in Sec.~\ref{subsec:Fterm}, 
the variation of $T$ and corresponding {\it F}-term are evaluated as
\begin{align}
\delta T&\simeq \frac{5i}{2a^2T_{\cal I}} \left[ 1-\frac{2iT_{\cal I}}{5}
\partial_T\ln ({\cal P})\right] ,
\nonumber\\
\left\langle\frac{F^T}{T-\bar{T}}\right\rangle &\simeq 
\frac{1}{aT_{\cal I}}m_{3/2} \left( \frac{5}{2}-i T_{\cal I}\partial_T \ln ({\cal P})\right),
\end{align}
at which the mass squared of $T$ is approximately the same in Eq.~(\ref{eq:Tmass}).
Throughout the discussion in Secs.~\ref{subsec:Fterm} and~\ref{subsec:uplift}, 
the complex structure modulus does not induce the SUSY-breaking at the uplifted 
minimum, if the uplifting sector is independent of the complex structure modulus. 
Thus, it can play a role of diluting the gravitino abundance as discussed in Sec.~\ref{sec:SUGRA}. 
In the next section, we focus on the moduli dynamics after the inflationary era.

\subsection{The dilution mechanism}
In this section, we repeat the dilution mechanism in Sec.~\ref{sec:SUGRA} by 
identifying $\Phi$ as the lightest complex structure modulus $U$. 
In a way similar to the step in Sec.~\ref{sec:SUGRA}, 
we define the true minimum of moduli fields as 
$T_0$ in Eq.~(\ref{eq:TRI}) for the K\"ahler modulus 
and $U_0$ in Eq.~(\ref{eq:URI}) for the lightest complex structure modulus, 
respectively. 
When the Hubble scale $H$ is comparable to the mass scale of moduli fields, $m_{T}$ and $m_U$, 
moduli fields roll down to the true minimum and dominate the energy density of 
the Universe as mentioned in Sec.~\ref{sec:SUGRA}. 
The scalar potential of our interest is 
\begin{align}
V_{\rm osc}=\frac{H^2(t)}{2} |T -T_1|^2 +\frac{m_{T}^2}{2} |T-T_0|^2 +
\frac{H^2(t)}{2} |U -U_1|^2 +\frac{m_{U}^2}{2} |U-U_0|^2,
\end{align}
where $U_1$ and $T_1$ denote the minimum induced by the Hubble parameter $H(t)$. 
Here, we employ the canonically normalized moduli fields by 
the same notation $U$ and $T$. 
In our model, the lightest complex structure modulus $U$ is lighter than the 
K\"ahler modulus $T$, i.e., $m_U < m_T$. 
Thus, $U$ oscillates subsequent to the oscillation of $T$, i.e., $t_{\rm osc}^U > t_{\rm osc}^T$, 
where the oscillating time of moduli fields $U$ and $T$ are defined as $t_{\rm osc}^U$ and 
$t_{\rm osc}^T$, respectively. 
As shown in Sec.~\ref{sec:SUGRA}, when we assume that the initial displacement of moduli fields as 
$\Delta T\simeq \Delta U={\cal O}(M_{\rm Pl})$, the oscillating energy densities of $U$ and $T$ at the time $t_{\rm osc}^U\simeq 1/m_U$ are almost the same as each other,
\begin{align}
\frac{1}{2}m_T^2T(t_{\rm osc}^U)^2\simeq \frac{1}{2}m_U^2 (\Delta U)^2. 
\end{align}
It implies that two moduli fields dominate the energy densities of Universe at the time $t_{\rm osc}^U$. 

Next, let us move onto the reheating process of both moduli fields. 
As analyzed in Eq.~(\ref{eq:Ttot}), the K\"ahler modulus dominantly decays into 
the gauge boson pairs, whereas the complex structure modulus does not 
appear in the gauge kinetic function at the tree-level. 
At the one-loop level, the gauge kinetic function has $U$-dependence through the threshold correction to 
the gauge coupling in the visible sector. 
We find that the one-loop gauge kinetic functions in the visible sector are brought into 
the following form,
\begin{align}
f_a =kT +\frac{1}{16\pi^2} \Delta_a(U), 
\label{eq:gknth}
\end{align}
where the threshold corrections $\Delta_a(U)$ are only calculated in the case of 
toroidal compactification~\cite{Dixon:1990pc,Lust:2003ky}. 
In this case, the authors of Ref.~\cite{Dixon:1990pc,Lust:2003ky} show that $\Delta_a(U)$ is 
written in terms of the Dedekind eta function and beta-function coefficient 
for the charged strings in the ${\cal N}=2$ sector. 
Thus, the decay width from $U$ into gauge boson pairs is suppressed by 
that of K\"ahler modulus. It is expected that the other possible decay channels 
are arisen from the kinetic terms of matter fields in the K\"ahler potential 
and Yukawa couplings in the superpotential. 
However, the axion associated with the lightest complex structure modulus $U$, $U_{\cal R}$, 
only appears in the kinetic terms of matter fields and Yukawa couplings 
through the non-perturbative effects due to the invariance of (discrete) shift symmetry. 
In this way, we assume that the lifetime of axion $U_{\cal R}$ becomes longer compared with $U_{\cal I}$ 
which appears in the tree-level K\"ahler and superpotentials, for simplicity.(See for example, Ref.~\cite{Kobayashi:2015aaa}.)
On the other hand, the decay temperatures of real and imaginary parts of $T$ are the same as shown in Eq.~(\ref{eq:Ttot}). 
As a result, the sizable decay channels of $U_{\cal R}$ are mainly determined 
through the gauge kinetic function~(\ref{eq:gknth}) and the total decay width of $U_{\cal R}$ is 
estimated as
\begin{align}
\Gamma_{\rm tot}^{U_{\cal R}} 
&=\sum_{a=1}^3 \frac{N^a}{128\pi} \left( \frac{\partial_U \Delta_a(U)}{16\pi^2} \right)^2
\left(\frac{g_a^2}{\sqrt{\langle K_{U\bar{U}}\rangle }}\right)^2 
\frac{m_{U_{\cal R}}^3}{M_{\rm Pl}^2}\nonumber\\
&\simeq  2.3\times 10^{-23} \left(\frac{\langle\Delta(U)\rangle}{20}\right)^2
\langle K^{U\bar{U}}\rangle\left( \frac{m_{U_{\cal R}}}{10^6\,{\rm GeV}}\right)^{3}\,{\rm GeV},
\label{eq:Utot}
\end{align}
with $\Delta(U)=\sum_a \partial_U \Delta_a(U)$, $N^a=\{8, 3, 1\}$ and $(g_a)^{2}\simeq 0.53$, 
which gives rise to a radiation-dominated Universe with a 
reheating temperature,
\begin{align}
T_{\rm reh}^{U_{\cal R}}&\simeq \left(\frac{\pi^2g_\ast (T_{\rm reh}^{U_{\cal R}})}{90}\right)^{-1/4} 
\sqrt{ M_{\rm Pl}\Gamma_{\rm tot}^{U_{\cal R}}} 
\nonumber\\
&\simeq  7.2 \left(\frac{\langle\Delta(U)\rangle}{20}\right)
\langle K^{U\bar{U}}\rangle^{1/2} 
\left( \frac{m_{U_{\cal R}}}{10^6\,{\rm GeV}}\right)^{3/2}\,{\rm MeV}
,
\end{align}
with $g_\ast (T_{\rm reh}^{U_{\cal R}}) \simeq 10.75$ being the effective number of degrees of freedom in the radiation-dominated Universe. 

It is argued that the lightest complex structure modulus $U_{\cal R}$ dominates the energy 
density of the Universe. Furthermore, this lightest modulus does not decay into the gravitino(s) 
due to the vanishing {\it F}-term of $U$. 
At the decay time of $U_{\cal R}$, $t_{\rm dec}^{U_{\cal R}}$, the produced entropy dilutes 
the gravitino (and LSP) yield given through the decay of $T$ at the time $t_{\rm dec}^T$. 
The dilution factor defined by the ratio between the entropy density of $U_{\cal R}$ ($s_{U_{\cal R}}$), 
and that of $T$ ($s_T$), at the time $t_{\rm dec}^{U_{\cal R}}$ is given by
\begin{align}
\Delta_{S} &\simeq \frac{s_{U_{\cal R}}(t_{\rm dec}^{U_{\cal R}})}{s_{T}(t_{\rm dec}^{U_{\cal R}})} 
\simeq \left(\frac{\Gamma_{\rm tot}^T}{\Gamma_{\rm tot}^{U_{\cal R}}}\right)^{1/2}
\nonumber\\
&\simeq 223\left(\frac{\langle K^{T\bar{T}}\rangle}{\langle K^{U\bar{U}}\rangle}\right)^{1/2}
\left( \frac{m_{T_{\cal R}}}{m_{U_{\cal R}}}\right)^{3/2}\left(\frac{|k|}{\langle\Delta(U)\rangle}\right),
\end{align}
from which the mass hierarchy between $m_{T_{\cal R}}$ and $m_{U_{\cal R}}$ 
leads the enhancement of dilution factor. 
In Secs.~\ref{subsec:Fterm} and~\ref{subsec:uplift}, 
we demonstrate the relation $m_{T_{\cal R}} \gg m_{U_{\cal R}}$ 
in the KKLT-and racetrack-type moduli stabilizations, respectively.

\begin{figure}[h]
\begin{minipage}{0.5\hsize}
\begin{center}
\includegraphics[width=0.8\linewidth]{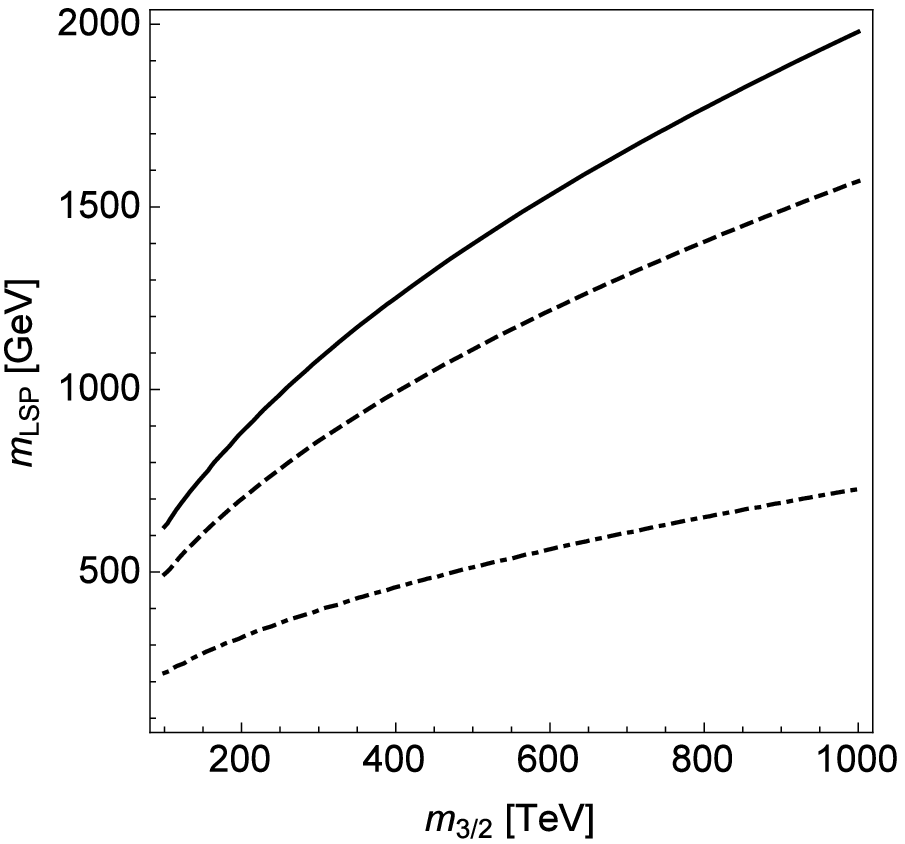}
\end{center}
\end{minipage}
\begin{minipage}{0.5\hsize}
\begin{center}
\includegraphics[width=0.8\linewidth]{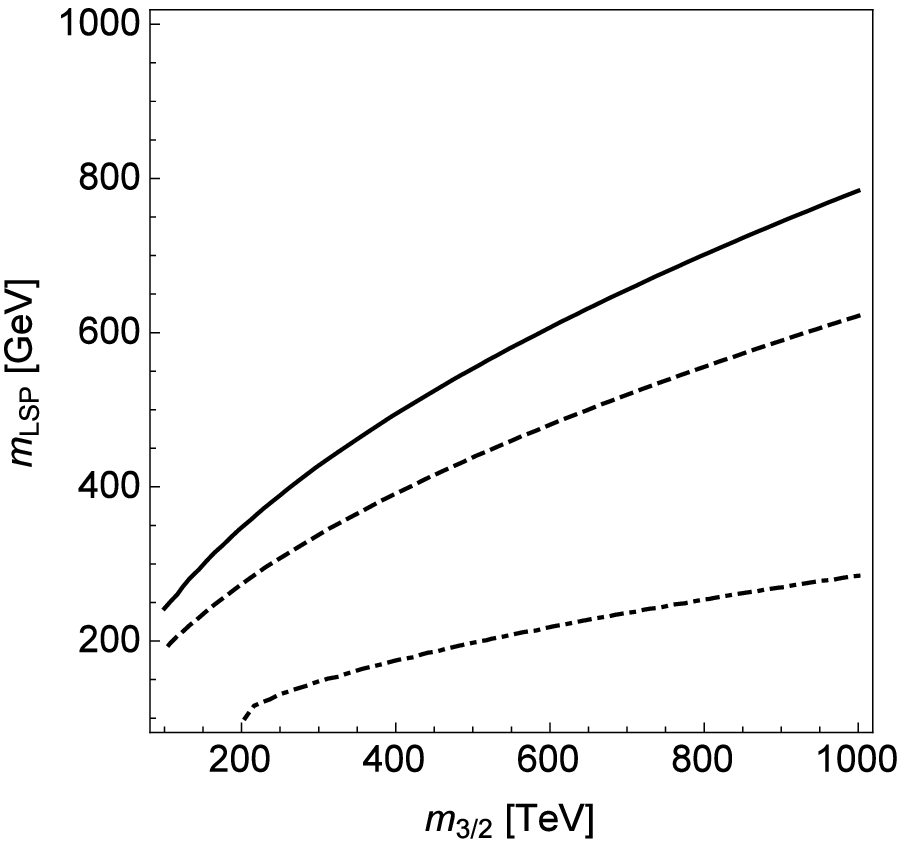}
\end{center}
\end{minipage}
\caption{The dark matter abundance on the ($m_{3/2}, m_{\rm LSP}$)-plane, 
where LSP is identified as the Wino-like (Higgsino-like) neturalino in the left (right) panel. 
The black solid, dashed, and dotdashed curves correspond to $\Omega_{\rm LSP}h^2=0.2$, 
$\Omega_{\rm LSP}h^2=0.1$, and $\Omega_{\rm LSP}h^2=0.01$, respectively.}
\label{fig}
\end{figure}

In the gravitino mass of $m_{3/2}\simeq {\cal O}(30)$ TeV, the BBN 
severely constrains the branching ratio $B_{3/2}$ in the modulus decay. 
However, the large dilution factor $\Delta_S$ reduces the gravitino yield at the 
time $t_{\rm dec}^{U_{\cal R}}$ to an acceptable level,
\begin{align}
&Y_{3/2}(t_{\rm dec}^{U_{\cal R}}) =\frac{n_{3/2}}{s_{U_{\cal R}}(t_{\rm dec}^{U_{\cal R}})}
=\frac{Y_{3/2}(t_{\rm dec}^T)}{\Delta_S}
\simeq B_{3/2}\frac{3T_{\rm reh}^{U_{\cal R}}}{2m_{T}}
\nonumber\\
&\simeq 2.2\times 10^{-12}\left(\frac{B_{3/2}}{0.01}\right) 
\left(\frac{\langle\Delta(U)\rangle}{20}\right)
\langle K^{U\bar{U}}\rangle^{1/2} 
\left( \frac{m_{U_{\cal R}}}{10^6\,{\rm GeV}}\right)^{3/2}
\left( \frac{5\times 10^7\,{\rm GeV}}{m_T}\right).
\label{eq:gradil}
\end{align}
Even if the gravitino is much larger than ${\cal O}(30)$ TeV, 
the LSP yield produced by gravitino decay is also diluted by the entropy production of $U_{\cal R}$,
\begin{align}
&Y_{\rm LSP}(t_{\rm dec}^{U_{\cal R}}) =\frac{Y_{\rm LSP}(t_{\rm dec}^T)}{\Delta_S}.
\end{align}
Since the large amount of gravitino is generated by the modulus decay 
in the high-scale SUSY-breaking scenario, 
the dark matter abundance is mainly dominated by the second term in Eq.~(\ref{eq:dmyield}),
\begin{align}
&\Omega_{\rm LSP}h^2\simeq 
m_{\rm LSP} Y_{\rm LSP}(t_{\rm dec}^{U_{\cal R}}) \frac{s_{\rm now}}{\rho_{\rm cr}} 
\nonumber\\
&\simeq 
m_{\rm LSP} 
\sqrt{\frac{45}{8\pi^2g_\ast (T_{\rm reh}^{U_{\cal R}})}} \frac{1}{M_{\rm Pl}T_{3/2}\langle \sigma_{\rm ann}v\rangle}
\frac{s_{\rm now}}{\rho_{\rm cr}}
\left(\frac{\Gamma_{\rm tot}^{U_{\cal R}}}{\Gamma_{\rm tot}^T}\right)^{1/2}
\nonumber\\
&\simeq 0.3
\left(\frac{80}{g_\ast (T_{3/2}) }\right)^{1/2}
\left(\frac{10^{6}\,{\rm GeV}}{m_{3/2}}\right)^{3/2}
\left( \frac{m_{\rm LSP}}{1\,{\rm TeV}}\right)^3
\left(\frac{10^{-3}\,{\rm GeV}^{-2}}{m_{\rm LSP}^2\langle \sigma_{\rm ann}v\rangle}\right)
\left(\frac{m_{U_{\cal R}}}{m_{T_{\cal R}}}\right)^{3/2}
\left(\frac{\langle\Delta(U)\rangle}{|k|}\right)
\left(\frac{\langle K^{U\bar{U}}\rangle}{\langle K^{T\bar{T}}\rangle}\right)^{1/2},
\end{align} 
which is inversely proportional to the gravitino mass. 
Note that two moduli masses ($m_{U_{\cal R}}$, $m_{T_{\cal R}}$) are both proportionally 
to the gravitino mass as discussed in Secs.~\ref{subsec:Fterm} and~\ref{subsec:uplift}. 
Thus, even if the gravitino mass is much heavier than ${\cal O}(30)$ TeV, 
the overabundance of the dark matter can be avoided in high-scale SUSY-breaking 
scenario. 
With the sample values of parameters achieved in the KKLT scenario, 
\begin{align} 
\langle \Delta (U)\rangle =10,\,\,\, |k|=1,\,\,\, 
\langle K^{U\bar{U}}\rangle \simeq \langle K^{T\bar{T}}\rangle \simeq 1,\,\,\,
m_{T_{\cal R}} \simeq 4\pi^2 m_{3/2},\,\,\, 
m_{U_{\cal R}} \simeq 2\pi m_{3/2},\,\,\,
\end{align} 
we find that the mass squared of gravitino and 
LSP is constrained to account for the observed dark matter abundance 
as illustrated in Fig.~\ref{fig}, where the left (right) panel considers the 
Wino-like (Higgsino-like) neutralino, respectively. 
Although we now focus on two LSP scenarios within the mass range 
$100\,{\rm GeV}< m_{\rm LSP} < 2000\,{\rm GeV}$ predicted by 
the mirage or anomaly mediations with gravitino mass 
$100\,{\rm TeV}< m_{3/2} < 1000\,{\rm TeV}$, 
it is straightforward to extend our analysis to other dark matter scenarios with 
different SUSY-breaking scale. 
 
As a result, in both the low-and high-scale SUSY breaking scenarios, 
the gravitino abundance is diluted by the huge entropy injection of the extra 
modulus field. 
In our scenario, the dilution mechanism is achieved by the existence of the lightest modulus field 
without breaking the SUSY. 
Although the kinetic mixing between $T$ and $U$ has the potential for causing the gravitino production, 
the ${\cal N}=2$ supersymmetric structure ensures the absence of kinetic mixing between 
moduli fields.

The dilution mechanism also has significant influences on the dark matter abundance 
of axion and the baryogenesis. 
Indeed, the entropy dilution allows us to treat the large decay constant of axion dark matter, 
compared with the usually considered axion dark matter~\cite{Kawasaki:1995vt}. 
Since the string theory naturally predicts the large axion decay constant, 
the axion could then become a plausible candidate of the cold dark matter. 
On the other hand, the late entropy dilution may face a problem to explain the present 
baryon asymmetry. 
However, the large initial baryon asymmetry before the moduli oscillation would 
lead to the nonnegligible baryon asymmetry, through 
the Affleck-Dine mechanism~\cite{Affleck:1984fy,Dine:1995kz}. 
We leave the detailed study of the baryon asymmetry to the future work.

\section{Conclusion}
\label{sec:con}
We have discussed the cosmological aspects of moduli and mirage mediations with 
an emphasis on the moduli-induced gravitino problem. 
Since the moduli fields gravitationally couples to the matter fields, they oscillate 
around their true minimum at a low temperature. 
The decay of moduli fields then generates the huge amount of gravitinos which 
can spoil the successful BBN in the low-scale SUSY-breaking scenario and cause the 
overabundance of LSP in the high-scale SUSY-breaking scenario, respectively. 
It is a common feature in the low-energy effective action originating from higher-dimensional theory, 
in particular, string theory. 
So far, there were several proposals to dilute the gravitinos via the thermal inflation~\cite{Lyth:1995hj,Lyth:1995ka}, 
Q-ball~\cite{Fujii:2001xp} and unstable domain-wall decays~\cite{Hattori:2015xla}, 
or the introduction of the axion sector~\cite{Nakamura:2008ey}. 

In this paper, we explore another dilution mechanism by taking into account the extra modulus field. 
In contrast to the KKLT-type moduli stabilization mechanism~\cite{Kachru:2003aw}, 
the extra modulus field has been 
stabilized at the supersymmetric minimum, as far as the uplifting sector does not depend on 
the extra modulus field. 
Thus, such extra field does not decay into the gravitinos at its decay. 
Furthermore, in the KKLT-type moduli stabilization, 
we have shown that this extra modulus field dominates the energy density of the Universe 
after the time when the volume modulus field dominates. 
This situation has been achieved when the extra modulus field is lighter than the volume modulus 
and at the same time, the decay time of extra modulus is later than that of volume modulus. 
Within the framework of type IIB (IIA) string theory, the lightest complex structure (K\"ahler) modulus 
corresponds to this plausible candidate. 
Thus, the mirage mediation is desirable scenario not only from the aspects of 
fine-tuning problem, 
but also from the cosmological aspects.

\section*{Acknowledgements}
The authors thank H. Abe for useful discussions. 
T.~K. was supported in part by
the Grant-in-Aid for Scientific Research No.~25400252 and No.~26247042 from the 
Ministry of Education,
Culture, Sports, Science and Technology (MEXT) in
Japan. 
H.~O. was supported in part by a Grant-in-Aid for JSPS Fellows 
No. 26-7296. 

\appendix

\section{Moduli-dependent uplifting scenario}
\label{app:uplift}
We comment on the case that the uplifting sector 
depends on lightest complex structure modulus $U$, 
in particular, the uplifting scenario with anti D-branes. 
As a consequence of anti D-branes, the uplifting potential 
is generically described by
\begin{align}
V_{\rm up} =e^{2K/3}{\cal P} (T-\bar{T}, U-\bar{U}),  
\end{align} 
where ${\cal P}$ is the function of moduli fields $T$ and $U$, in general. 
In contrast to the discussion in Sec.~\ref{subsec:uplift}, 
the minimum of the lightest complex structure modulus $U$ can be deviated 
from the supersymmetric minimum, $K_U=0$. 
Along the same step outlined in the Sec.~\ref{subsec:Fterm}, 
the variations of $T$ and $U$ are evaluated in terms of
\begin{align}
&D_TW =W_{TT}|_{\rm ref} \delta T +(K_T W_T)|_{\rm ref}\delta T +( K_{T\bar{T}} W)|_{\rm ref} 
\left( \delta \bar{T}-\delta T\right),
\nonumber\\
&D_UW= (K_{UU}  W)|_{\rm ref}\delta U +(K_{U\bar{U}}  W)|_{\rm ref}\delta \bar{U}.
\end{align}
At quadratic order of $\delta T$ and $\delta U$ in the scalar potential, 
we obtain the variations of $T$ and $U$ 
\begin{align}
\delta T&\simeq \frac{5i}{2a^2T_{\cal I}} \left[ 1-\frac{2iT_{\cal I}}{5}
\partial_T\ln ({\cal P})\right] 
+\frac{4\partial_U\ln ({\cal P})
\left(\frac{\partial_T\partial_U P}{P}+\frac{i}{T_{\cal I}}\partial_U\ln ({\cal P})\right)}{a^2\left[ \frac{K^{U\bar{U}}}{6}(K_{U_{\cal I}U_{\cal I}})^2 -4\left(\frac{2}{3}K_{UU} 
+\frac{\partial_U\partial_U P}{P}\right)\right]},
\nonumber\\
\delta U&\simeq \frac{2\partial_U \ln ({\cal P})}
{\frac{K^{U\bar{U}}(K_{U_{\cal I}U_{\cal I}})^2}{6} -4\left(\frac{2}{3}K_{UU} 
+\frac{\partial_U\partial_U P}{P}\right)}.
\end{align}
Since there is no kinetic mixing between $U$ and $T$, 
the mass squared of canonically normalized moduli fields are evaluated at the 
obtained minimum, $T=T_{\rm ref} +\delta T$ and $U=U_{\rm ref}+\delta U$, 
\begin{align}
m_{U_{\cal R}}^2&\simeq (2\pi)^2 \left( U_{\cal I} +\frac{1}{\pi}\right)^2 m_{3/2}^2 
+4\pi \left(U_{\cal I} +\frac{1}{\pi}\right) m_{3/2}^2,
\nonumber\\
m_{T_{\cal R}}^2 &\simeq 2(aT_{\cal I})^2 m_{3/2}^2,
\end{align}
for real part of moduli fields. By contrast, 
the mass matrices of imaginary parts are given in the basis $(U_{\cal I}, T_{\cal I})$,
\begin{align}
m_{\cal I}^2&=
\begin{pmatrix}
\frac{\partial_{U_{\cal I}}\partial_{U_{\cal I}} V}{2K_{U\bar{U}}} 
& \frac{\partial_{U_{\cal I}}\partial_{T_{\cal I}} V}{2\sqrt{K_{U\bar{U}}} \sqrt{K_{T\bar{T}}}}\\
\frac{\partial_{U_{\cal I}}\partial_{T_{\cal I}} V}{2\sqrt{K_{U\bar{U}}} \sqrt{K_{T\bar{T}}}}
& \frac{\partial_{T_{\cal I}}\partial_{T_{\cal I}} V}{2K_{T\bar{T}}}\\
\end{pmatrix}
\nonumber\\
&\simeq 
\begin{pmatrix}
(2\pi)^2 U_{\cal I}^2 \left( 1+\frac{3}{2K_{U_{\cal I}U_{\cal I}}} 
\frac{\partial_{U_{\cal I}}\partial_{U_{\cal I}}P}{P}\right)
& \frac{3}{2\sqrt{K_{U\bar{U}}} \sqrt{K_{T\bar{T}}}}
\biggl[ \frac{\partial_{T_{\cal I}}\partial_{U_{\cal I}} P  }{P} +\frac{2}{3}K_{T_{\cal I}}\partial_{U_{\cal I}} 
\ln (P)\biggl]\\
\frac{3}{2\sqrt{K_{U\bar{U}}} \sqrt{K_{T\bar{T}}}}
\biggl[ \frac{\partial_{T_{\cal I}}\partial_{U_{\cal I}} P}{P} +\frac{2}{3}K_{T_{\cal I}}\partial_{U_{\cal I}} 
\ln (P)\biggl]
&2(aT_{\cal I})^2 
\end{pmatrix}
m_{3/2}^2,
\end{align}
which shows that when the derivatives of ${\cal P}$ with respect to the moduli fields are of 
${\cal O}({\cal P})$, the mass squared of moduli fields can be positive in the limit of 
$aT_{\cal I}\gg 1$.
According to it, {\it F}-terms of moduli fields are 
\begin{align}
\left\langle\frac{F^T}{T-\bar{T}}\right\rangle &\simeq 
\frac{1}{aT_{\cal I}}m_{3/2} 
\Biggl[\left( \frac{5}{2}-i T_{\cal I}\partial_T \ln ({\cal P})\right) 
+\frac{24(\partial_U \ln {\cal P}) \left( (\partial_T\partial_U P)/P +i\frac{1}{T_{\cal I}} \partial_U \ln ({\cal P})
\right)}{T_{\cal I} \left( K^{U\bar{U}} (K_{U_{\cal I}U_{\cal I}})^2 -16K_{UU}
(\partial_U\partial_U P)/P\right)}\Biggl],
\nonumber\\
\left\langle\frac{F^U}{U-\bar{U}}\right\rangle &\simeq 
12\pi 
\frac{-i\partial_U \ln {\cal P}}{ K^{U\bar{U}}(K_{U_{\cal I}U_{\cal I}})^2-16K_{UU}
-24(\partial_U\partial_U P)/P}
m_{3/2}.
\end{align}
It turns out that the complex structure modulus has a sizable {\it F}-term, 
although it depends on the details of the uplifting sector.


\begin{thebibliography}{99}


 
\bibitem{Choi:2004sx}
  K.~Choi, A.~Falkowski, H.~P.~Nilles, M.~Olechowski and S.~Pokorski,
  JHEP {\bf 0411} (2004) 076
  doi:10.1088/1126-6708/2004/11/076
  [hep-th/0411066].
  
\bibitem{Choi:2005ge}
  K.~Choi, A.~Falkowski, H.~P.~Nilles and M.~Olechowski,
  Nucl.\ Phys.\ B {\bf 718} (2005) 113
  doi:10.1016/j.nuclphysb.2005.04.032
  [hep-th/0503216].
  


\bibitem{Choi:2005uz}
  K.~Choi, K.~S.~Jeong and K.~i.~Okumura,
  JHEP {\bf 0509} (2005) 039
  doi:10.1088/1126-6708/2005/09/039
  [hep-ph/0504037].
  
\bibitem{Endo:2005uy}
  M.~Endo, M.~Yamaguchi and K.~Yoshioka,
  Phys.\ Rev.\ D {\bf 72} (2005) 015004
  doi:10.1103/PhysRevD.72.015004
  [hep-ph/0504036].
  
  
\bibitem{Kaplunovsky:1993rd}
  V.~S.~Kaplunovsky and J.~Louis,
  Phys.\ Lett.\ B {\bf 306} (1993) 269
  doi:10.1016/0370-2693(93)90078-V
  [hep-th/9303040].
  
  
\bibitem{Brignole:1993dj}
  A.~Brignole, L.~E.~Ibanez and C.~Munoz,
  Nucl.\ Phys.\ B {\bf 422} (1994) 125
   [Nucl.\ Phys.\ B {\bf 436} (1995) 747]
  doi:10.1016/0550-3213(94)00068-9
  [hep-ph/9308271].
  
  
\bibitem{Kobayashi:1994eh} 
  T.~Kobayashi, D.~Suematsu, K.~Yamada and Y.~Yamagishi,
  Phys.\ Lett.\ B {\bf 348}, 402 (1995)
  doi:10.1016/0370-2693(95)00194-P
  [hep-ph/9408322].
  
\bibitem{Ibanez:1998rf} 
  L.~E.~Ibanez, C.~Munoz and S.~Rigolin,
  Nucl.\ Phys.\ B {\bf 553}, 43 (1999)
  doi:10.1016/S0550-3213(99)00264-3
  [hep-ph/9812397].


 
\bibitem{Randall:1998uk}
  L.~Randall and R.~Sundrum,
  Nucl.\ Phys.\ B {\bf 557} (1999) 79
  doi:10.1016/S0550-3213(99)00359-4
  [hep-th/9810155].
  
\bibitem{Giudice:1998xp}
  G.~F.~Giudice, M.~A.~Luty, H.~Murayama and R.~Rattazzi,
  JHEP {\bf 9812} (1998) 027
  doi:10.1088/1126-6708/1998/12/027
  [hep-ph/9810442].
  
\bibitem{Kachru:2003aw}
  S.~Kachru, R.~Kallosh, A.~D.~Linde and S.~P.~Trivedi,
  Phys.\ Rev.\ D {\bf 68} (2003) 046005
  doi:10.1103/PhysRevD.68.046005
  [hep-th/0301240].


\bibitem{Choi:2005hd} 
  K.~Choi, K.~S.~Jeong, T.~Kobayashi and K.~i.~Okumura,
  Phys.\ Lett.\ B {\bf 633}, 355 (2006)
  doi:10.1016/j.physletb.2005.11.078
  [hep-ph/0508029].
  
 
 
  
  
\bibitem{Kitano:2005wc}
  R.~Kitano and Y.~Nomura,
  Phys.\ Lett.\ B {\bf 631} (2005) 58
  doi:10.1016/j.physletb.2005.10.003
  [hep-ph/0509039].
  
  
  
\bibitem{Choi:2006xb}
  K.~Choi, K.~S.~Jeong, T.~Kobayashi and K.~i.~Okumura,
  Phys.\ Rev.\ D {\bf 75} (2007) 095012
  doi:10.1103/PhysRevD.75.095012
  [hep-ph/0612258].
  
\bibitem{Kobayashi:2012ee}
  T.~Kobayashi, H.~Makino, K.~i.~Okumura, T.~Shimomura and T.~Takahashi,
  JHEP {\bf 1301} (2013) 081
  doi:10.1007/JHEP01(2013)081
  [arXiv:1204.3561 [hep-ph]].
   
  
 
\bibitem{Hagimoto:2015tua} 
  K.~Hagimoto, T.~Kobayashi, H.~Makino, K.~i.~Okumura and T.~Shimomura,
  JHEP {\bf 1602}, 089 (2016)
  doi:10.1007/JHEP02(2016)089
  [arXiv:1509.05327 [hep-ph]].
 
  
\bibitem{Asano:2012sv}
  M.~Asano and T.~Higaki,
  Phys.\ Rev.\ D {\bf 86} (2012) 035020
  doi:10.1103/PhysRevD.86.035020
  [arXiv:1204.0508 [hep-ph]].
  
 
  
\bibitem{Endo:2006zj}
  M.~Endo, K.~Hamaguchi and F.~Takahashi,
  Phys.\ Rev.\ Lett.\  {\bf 96} (2006) 211301
  doi:10.1103/PhysRevLett.96.211301
  [hep-ph/0602061].
  
\bibitem{Nakamura:2006uc}
  S.~Nakamura and M.~Yamaguchi,
  Phys.\ Lett.\ B {\bf 638} (2006) 389
  doi:10.1016/j.physletb.2006.05.078
  [hep-ph/0602081].
  
\bibitem{Dine:2006ii}
  M.~Dine, R.~Kitano, A.~Morisse and Y.~Shirman,
  Phys.\ Rev.\ D {\bf 73} (2006) 123518
  doi:10.1103/PhysRevD.73.123518
  [hep-ph/0604140].
  
\bibitem{Lyth:1995hj}
  D.~H.~Lyth and E.~D.~Stewart,
  Phys.\ Rev.\ Lett.\  {\bf 75} (1995) 201
  doi:10.1103/PhysRevLett.75.201
  [hep-ph/9502417].
  
\bibitem{Lyth:1995ka}
  D.~H.~Lyth and E.~D.~Stewart,
  Phys.\ Rev.\ D {\bf 53} (1996) 1784
  doi:10.1103/PhysRevD.53.1784
  [hep-ph/9510204].
  
\bibitem{Fujii:2001xp}
  M.~Fujii and K.~Hamaguchi,
  Phys.\ Lett.\ B {\bf 525} (2002) 143
  doi:10.1016/S0370-2693(01)01412-5
  [hep-ph/0110072].
  
\bibitem{Hattori:2015xla}
  H.~Hattori, T.~Kobayashi, N.~Omoto and O.~Seto,
  Phys.\ Rev.\ D {\bf 92} (2015) 10,  103518
  doi:10.1103/PhysRevD.92.103518
  [arXiv:1510.03595 [hep-ph]].
  
\bibitem{Nakamura:2008ey}
  S.~Nakamura, K.~i.~Okumura and M.~Yamaguchi,
  Phys.\ Rev.\ D {\bf 77} (2008) 115027
  doi:10.1103/PhysRevD.77.115027
  [arXiv:0803.3725 [hep-ph]].
  
\bibitem{Linde:1996cx}
  A.~D.~Linde,
  Phys.\ Rev.\ D {\bf 53} (1996) 4129
  doi:10.1103/PhysRevD.53.R4129
  [hep-th/9601083].
  
\bibitem{Nakayama:2011wqa}
  K.~Nakayama, F.~Takahashi and T.~T.~Yanagida,
  Phys.\ Rev.\ D {\bf 84} (2011) 123523
  doi:10.1103/PhysRevD.84.123523
  [arXiv:1109.2073 [hep-ph]].
  
  
\bibitem{Balasubramanian:2005zx}
  V.~Balasubramanian, P.~Berglund, J.~P.~Conlon and F.~Quevedo,
  JHEP {\bf 0503} (2005) 007
  doi:10.1088/1126-6708/2005/03/007
  [hep-th/0502058].
  
\bibitem{Conlon:2005ki}
  J.~P.~Conlon, F.~Quevedo and K.~Suruliz,
  JHEP {\bf 0508} (2005) 007
  doi:10.1088/1126-6708/2005/08/007
  [hep-th/0505076].
  
\bibitem{Giddings:2001yu}
  S.~B.~Giddings, S.~Kachru and J.~Polchinski,
  Phys.\ Rev.\ D {\bf 66} (2002) 106006
  doi:10.1103/PhysRevD.66.106006
  [hep-th/0105097].
  
  
\bibitem{Gukov:1999ya}
  S.~Gukov, C.~Vafa and E.~Witten,
  Nucl.\ Phys.\ B {\bf 584} (2000) 69
   [Nucl.\ Phys.\ B {\bf 608} (2001) 477]
  doi:10.1016/S0550-3213(00)00373-4
  [hep-th/9906070].
  
\bibitem{Krasnikov:1987jj}
  N.~V.~Krasnikov,
  Phys.\ Lett.\ B {\bf 193} (1987) 37.
  doi:10.1016/0370-2693(87)90452-7
  
\bibitem{Dixon:1990ds} 
  L.~J.~Dixon,
  SLAC-PUB-5229, C90-01-03;
  
  
\bibitem{Lebedev:2006qq} 
  O.~Lebedev, H.~P.~Nilles and M.~Ratz,
  Phys.\ Lett.\ B {\bf 636}, 126 (2006)
  [hep-th/0603047].
  
  
\bibitem{Dudas:2006gr}
  E.~Dudas, C.~Papineau and S.~Pokorski,
  JHEP {\bf 0702} (2007) 028
  doi:10.1088/1126-6708/2007/02/028
  [hep-th/0610297].
  
  
\bibitem{Abe:2006xp}
  H.~Abe, T.~Higaki, T.~Kobayashi and Y.~Omura,
  Phys.\ Rev.\ D {\bf 75} (2007) 025019
  doi:10.1103/PhysRevD.75.025019
  [hep-th/0611024].
  
\bibitem{Kallosh:2006dv}
  R.~Kallosh and A.~D.~Linde,
  JHEP {\bf 0702} (2007) 002
  doi:10.1088/1126-6708/2007/02/002
  [hep-th/0611183].
  
\bibitem{Abe:2007yb}
  H.~Abe, T.~Higaki and T.~Kobayashi,
  Phys.\ Rev.\ D {\bf 76} (2007) 105003
  doi:10.1103/PhysRevD.76.105003
  [arXiv:0707.2671 [hep-th]].
  
\bibitem{Moroi:1999zb}
  T.~Moroi and L.~Randall,
  Nucl.\ Phys.\ B {\bf 570} (2000) 455
  doi:10.1016/S0550-3213(99)00748-8
  [hep-ph/9906527].
  
\bibitem{Kawasaki:2004qu}
  M.~Kawasaki, K.~Kohri and T.~Moroi,
  Phys.\ Rev.\ D {\bf 71} (2005) 083502
  doi:10.1103/PhysRevD.71.083502
  [astro-ph/0408426].
  
\bibitem{Kawasaki:2004yh}
  M.~Kawasaki, K.~Kohri and T.~Moroi,
  Phys.\ Lett.\ B {\bf 625} (2005) 7
  doi:10.1016/j.physletb.2005.08.045
  [astro-ph/0402490].
  
\bibitem{Kohri:2005wn}
  K.~Kohri, T.~Moroi and A.~Yotsuyanagi,
  Phys.\ Rev.\ D {\bf 73} (2006) 123511
  doi:10.1103/PhysRevD.73.123511
  [hep-ph/0507245].
  
  
\bibitem{Olive:1989jg}
  K.~A.~Olive and M.~Srednicki,
  Phys.\ Lett.\ B {\bf 230} (1989) 78.
  doi:10.1016/0370-2693(89)91656-0
  
\bibitem{Ade:2013zuv}
  P.~A.~R.~Ade {\it et al.} [Planck Collaboration],
  Astron.\ Astrophys.\  {\bf 571} (2014) A16
  doi:10.1051/0004-6361/201321591
  [arXiv:1303.5076 [astro-ph.CO]].
  
\bibitem{Ade:2015lrj}
  P.~A.~R.~Ade {\it et al.} [Planck Collaboration],
  arXiv:1502.02114 [astro-ph.CO].
  
  
\bibitem{Conlon:2006tq}
  J.~P.~Conlon,
  JHEP {\bf 0605} (2006) 078
  doi:10.1088/1126-6708/2006/05/078
  [hep-th/0602233].
  
  
\bibitem{Kawasaki:2006gs}
  M.~Kawasaki, F.~Takahashi and T.~T.~Yanagida,
  Phys.\ Lett.\ B {\bf 638} (2006) 8
  doi:10.1016/j.physletb.2006.05.037
  [hep-ph/0603265].
  
\bibitem{Hori:2000kt}
  K.~Hori and C.~Vafa,
  hep-th/0002222.
  

\bibitem{Dixon:1990pc} 
  L.~J.~Dixon, V.~Kaplunovsky and J.~Louis,
  Nucl.\ Phys.\ B {\bf 355}, 649 (1991).
  doi:10.1016/0550-3213(91)90490-O



\bibitem{Lust:2003ky}
  D.~Lust and S.~Stieberger,
  Fortsch.\ Phys.\  {\bf 55} (2007) 427
  doi:10.1002/prop.200310335
  [hep-th/0302221].  
  
  
\bibitem{Hosono:1993qy}
  S.~Hosono, A.~Klemm, S.~Theisen and S.~T.~Yau,
  Commun.\ Math.\ Phys.\  {\bf 167} (1995) 301
  doi:10.1007/BF02100589
  [hep-th/9308122].
  
\bibitem{Hosono:1994ax}
  S.~Hosono, A.~Klemm, S.~Theisen and S.~T.~Yau,
  Nucl.\ Phys.\ B {\bf 433} (1995) 501
  doi:10.1016/0550-3213(94)00440-P
  [hep-th/9406055].
  
\bibitem{Intriligator:2006dd}
  K.~A.~Intriligator, N.~Seiberg and D.~Shih,
  JHEP {\bf 0604} (2006) 021
  doi:10.1088/1126-6708/2006/04/021
  [hep-th/0602239].
 
\bibitem{Kobayashi:2015aaa}
  T.~Kobayashi, A.~Oikawa and H.~Otsuka,
  arXiv:1510.08768 [hep-ph].
  

\bibitem{Kawasaki:1995vt}
  M.~Kawasaki, T.~Moroi and T.~Yanagida,
  Phys.\ Lett.\ B {\bf 383} (1996) 313
  doi:10.1016/0370-2693(96)00743-5
  [hep-ph/9510461].
  
\bibitem{Affleck:1984fy}
  I.~Affleck and M.~Dine,
  Nucl.\ Phys.\ B {\bf 249} (1985) 361.
  doi:10.1016/0550-3213(85)90021-5
  
\bibitem{Dine:1995kz}
  M.~Dine, L.~Randall and S.~D.~Thomas,
  Nucl.\ Phys.\ B {\bf 458} (1996) 291
  doi:10.1016/0550-3213(95)00538-2
  [hep-ph/9507453].



  
  
\end{thebibliography}

\end{document}